\documentclass[12pt,twoside]{article}
\usepackage{fleqn,espcrc1,epsf}

\newcommand{\AmS}{{\protect\the\textfont2
  A\kern-.1667em\lower.5ex\hbox{M}\kern-.125emS}}

\newcommand{\Gammat}{{\mit\Gamma_{\rm t}}}
\newcommand{\Gammas}{{\mit\Gamma_{\rm s}}}
\newcommand{\Gamman}{{\mit\Gamma_{\rm n}}}
\newcommand{\Dn}{{D_{\rm n}}}

\newcommand{\Nout}{{N_{\rm out}}}
\newcommand{\Iout}{{I_{\rm out}}}
\newcommand{\Cmass}{{C_{\rm mass}}}
\newcommand{\I}{{ {\rm I} }}
\newcommand{\II}{{ {\rm II} }}
\newcommand{\IItoI}{{ {\rm II}\hbox{\hbox{\rm -}\kern-0.1em}{\rm I} }}
\newcommand{\ston}{{ {\rm s}\hbox{\hbox{\rm -}\kern-0.1em}{\rm n} }}

\newcommand{\rmn}{{ {\rm n} }}
\newcommand{\Nband}{{N_{\rm band}}}
\newcommand{\FG}{{\rm FG}}
\newcommand{\mGamma}{{\mit\Gamma}}
\newcommand{\Eone}{{\rm E1}}
\newcommand{\omegas}{{\omega_{\rm s}}}

\title{Theoretical study of the decay-out spin of superdeformed bands
       in the Dy and Hg regions}

\author{Y.~R.~Shimizu\address{
     Department of Physics, Faculty of Sciences, Kyushu University,
     Fukuoka 812-8581, Japan},
        M.~Matsuo\address{
     Graduate School of Science and Technology, Niigata University,
     Niigata 950-2101, Japan}
        and   
        K.~Yoshida\address{
     Institute of Natural Science, Nara University, Nara 631-8502, Japan}
}
       
\begin{document}

\maketitle

\begin{abstract}
Decay of the superdeformed bands have been studied
mainly concentrating upon the decay-out spin,
which is sensitive to the tunneling probability between
the super- and normal-deformed wells.
Although the basic features are well understood by the calculations,
it is difficult to precisely reproduce
the decay-out spins in some cases.
Comparison of the systematic calculations
with experimental data reveals that
values of the calculated decay-out spins scatter more broadly
around the average value in both
the $A \approx$ 150 and 190 regions, which reflects the variety
of calculated tunneling probability in each band.
\end{abstract}

\section{MODEL OF THE DECAY OF SUPERDEFORMED BANDS}

\begin{figure}[t]
\centerline{
\epsfxsize=40mm\epsffile{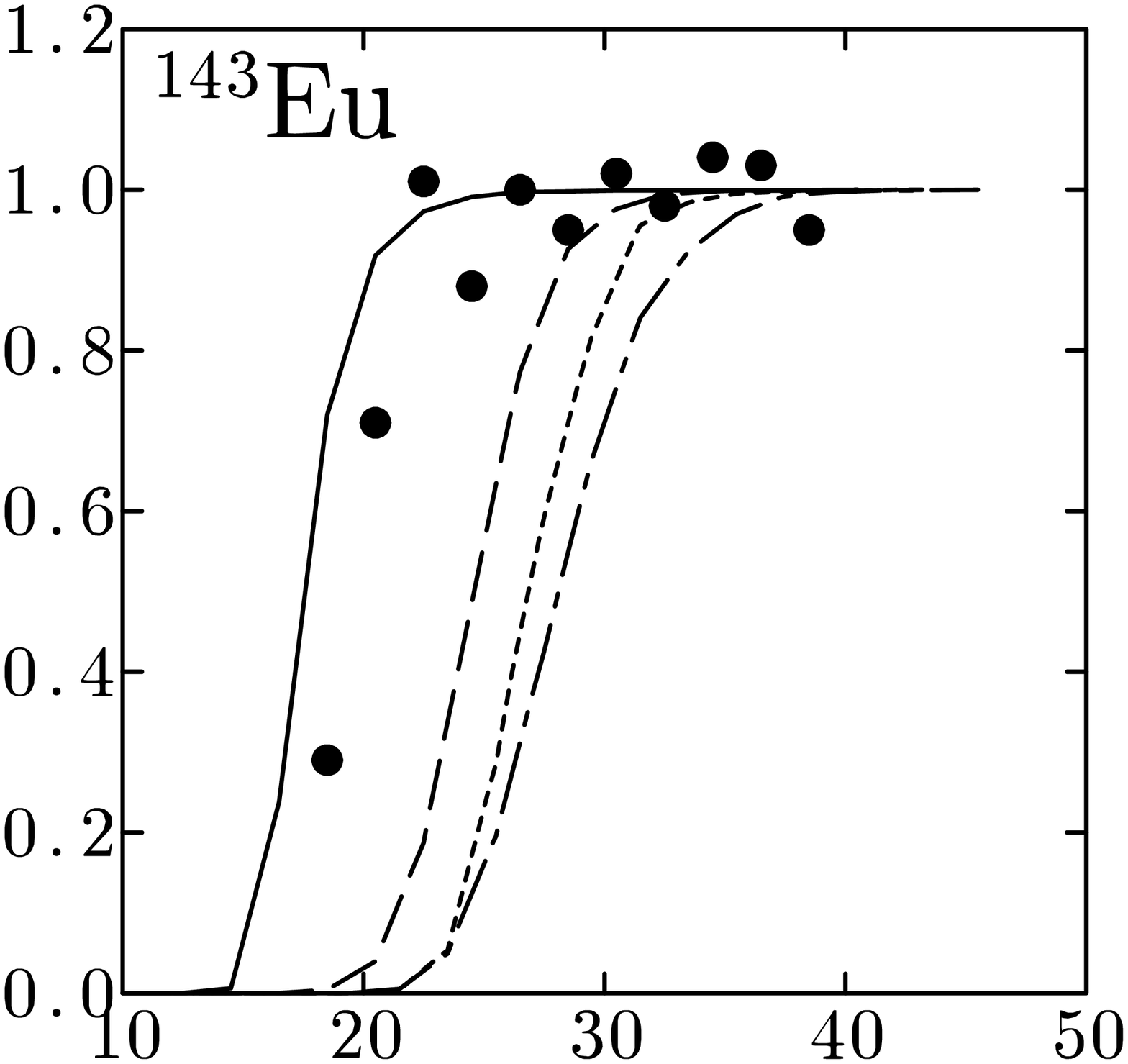} \hspace{-4mm}
\epsfxsize=40mm\epsffile{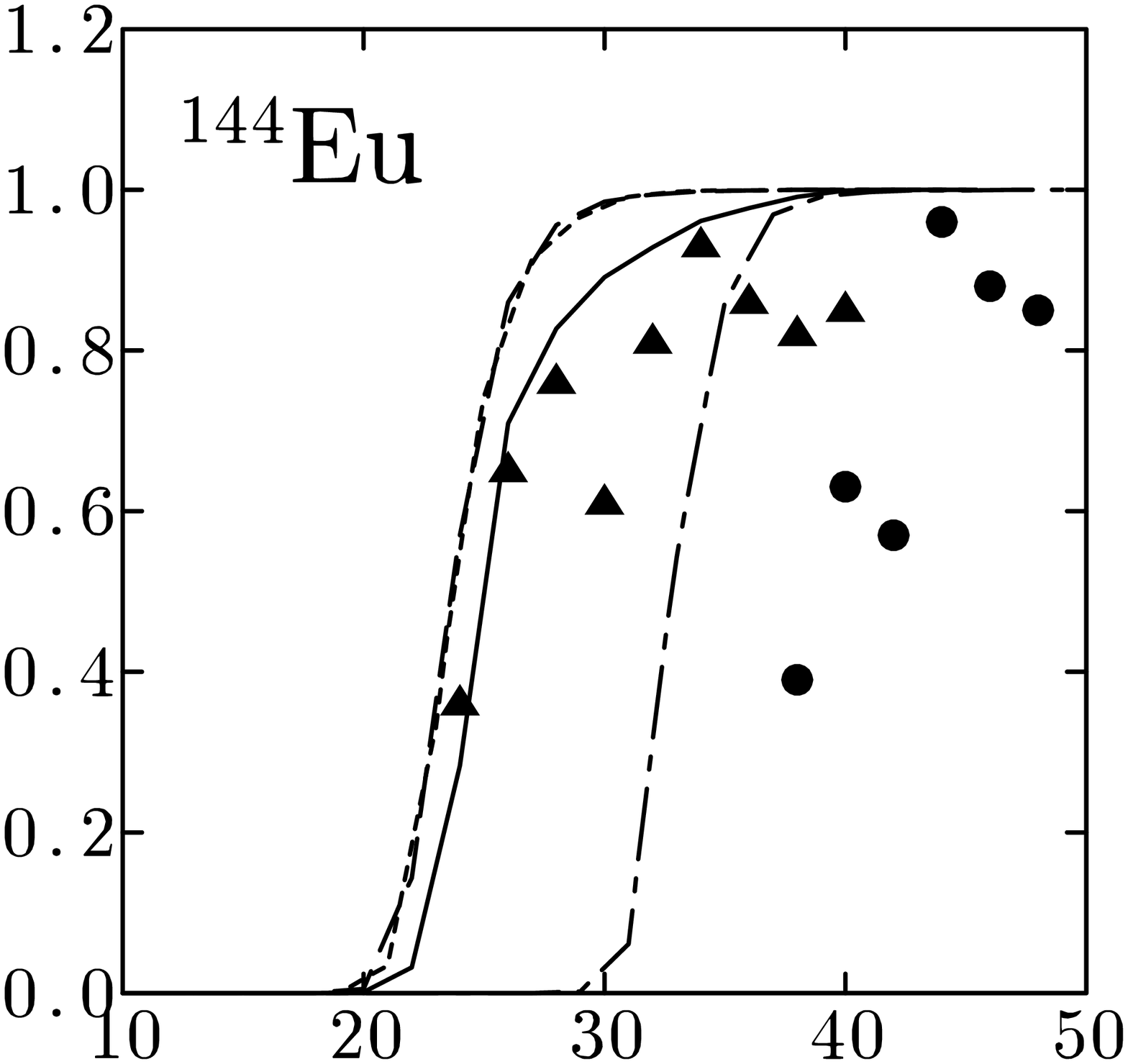} \hspace{-4mm}
\epsfxsize=40mm\epsffile{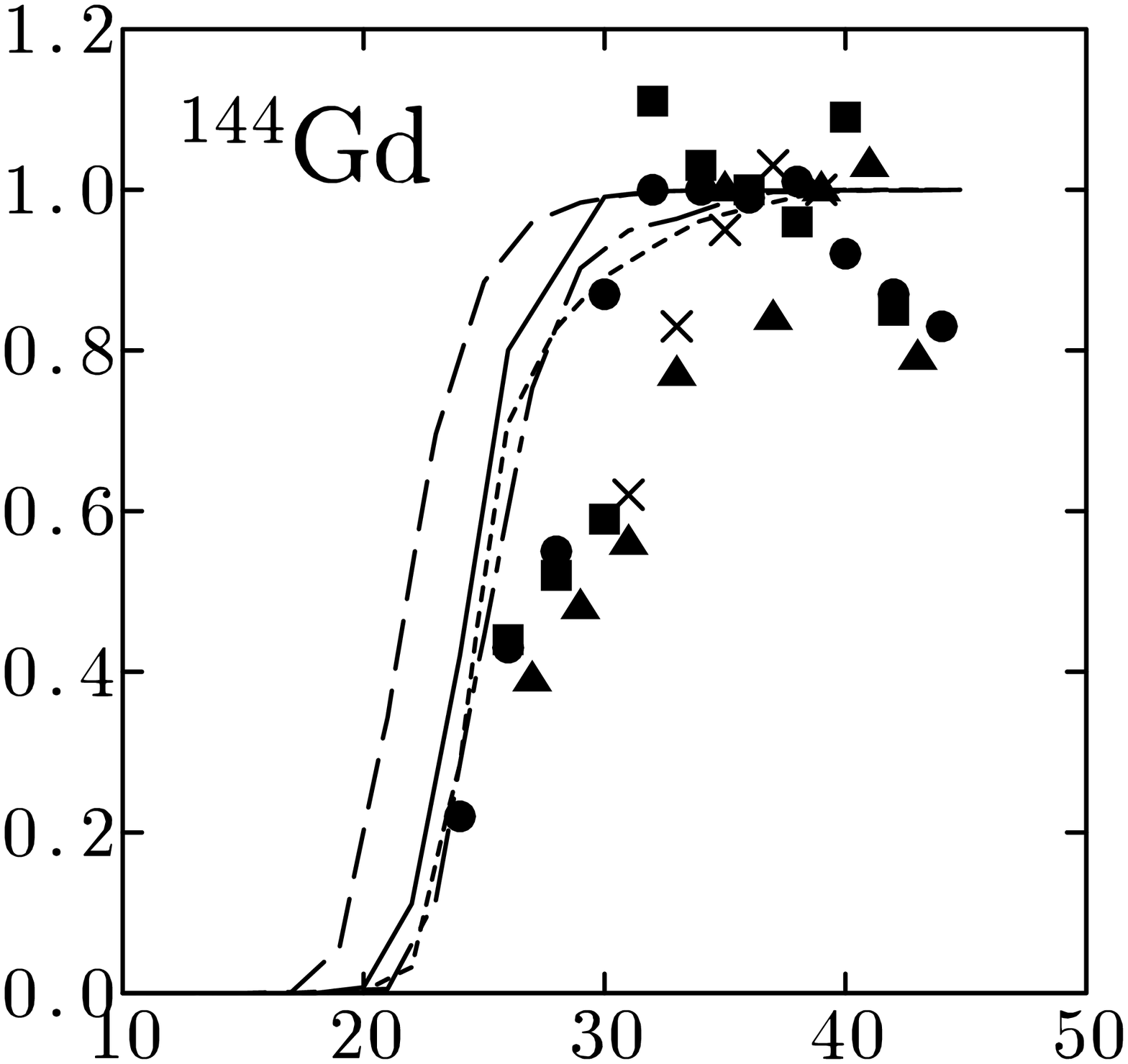} \hspace{-4mm}
\epsfxsize=40mm\epsffile{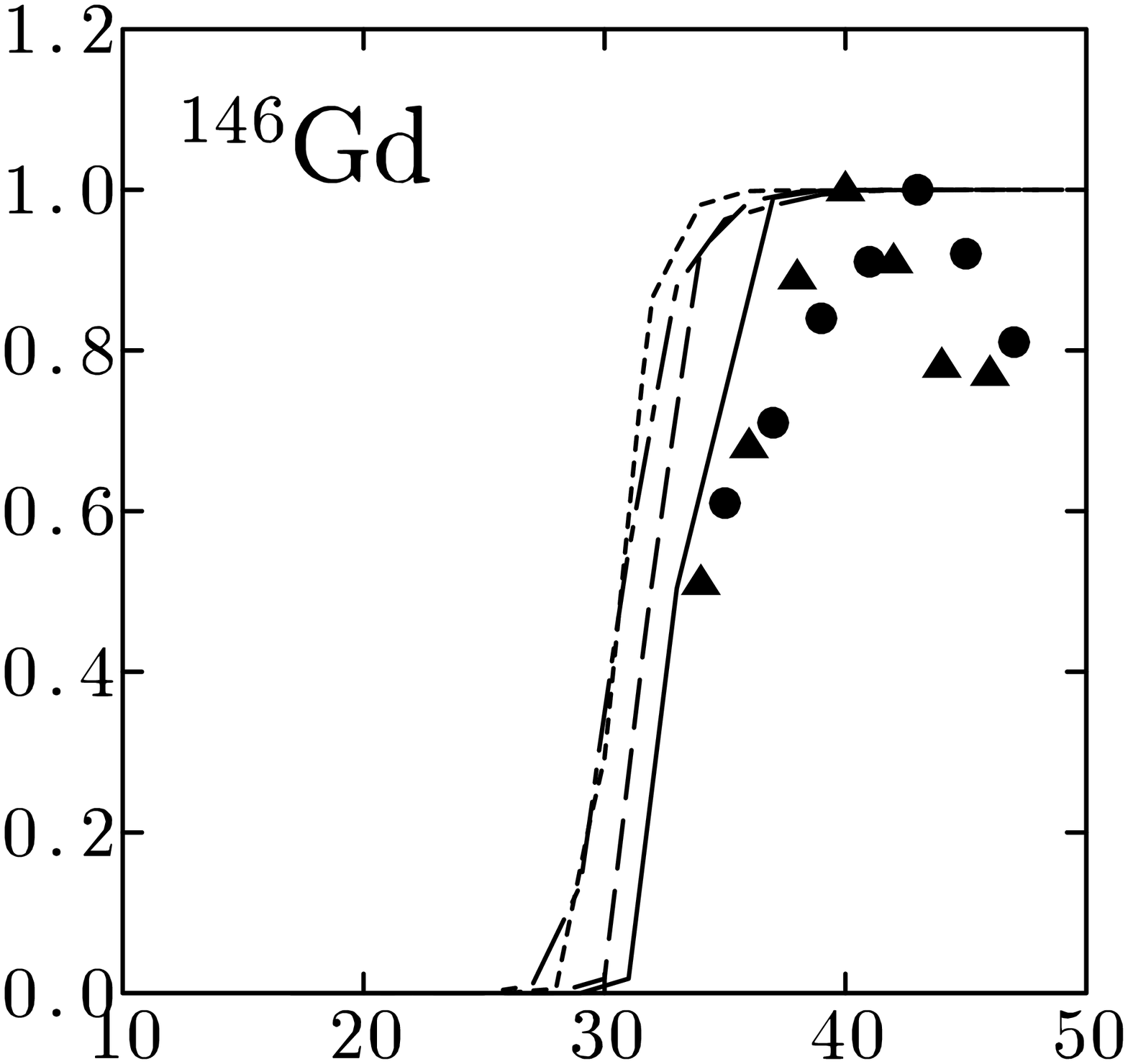} \hspace{-4mm}
}
\vspace{-2mm}
\centerline{
\epsfxsize=40mm\epsffile{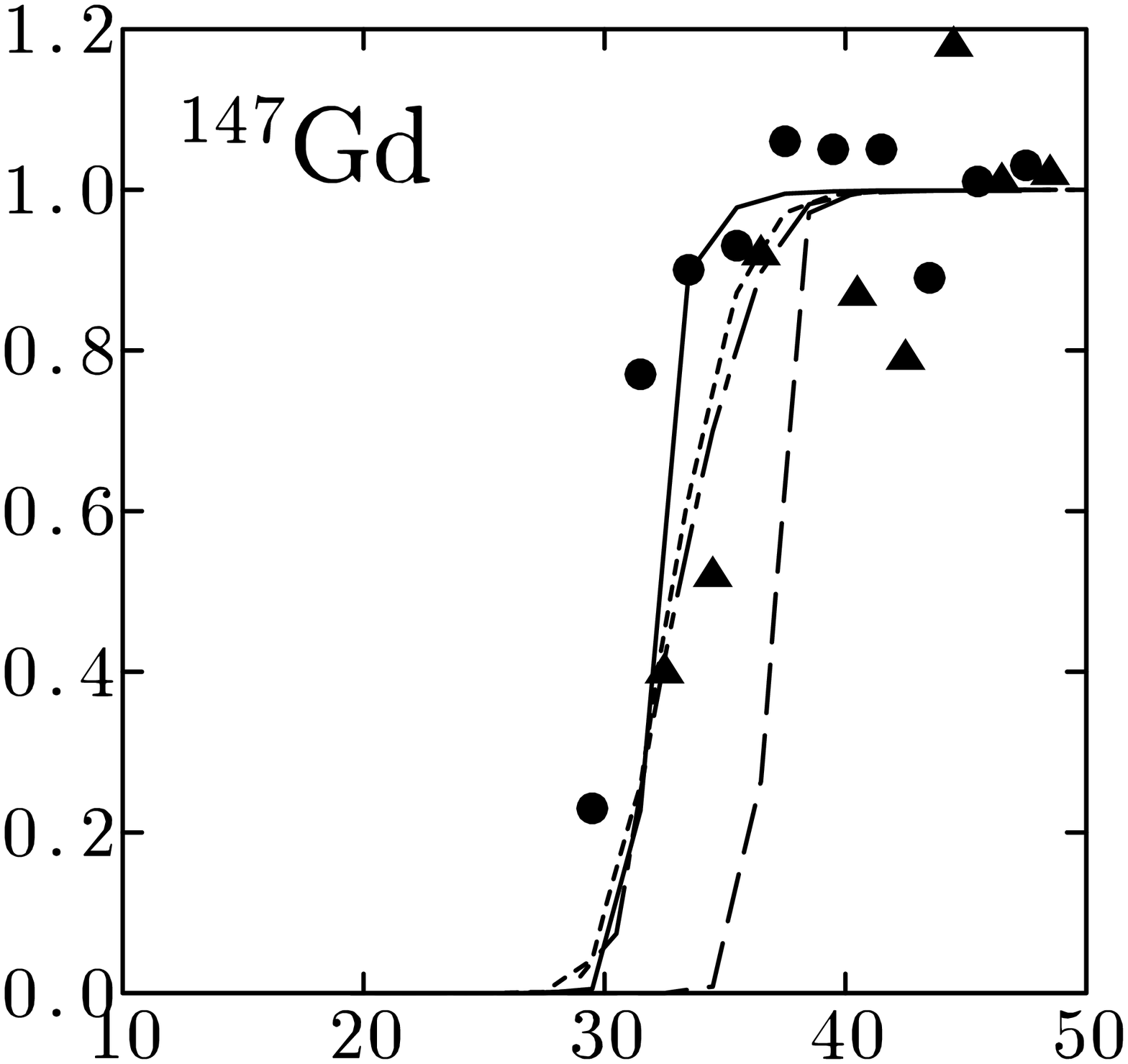} \hspace{-4mm}
\epsfxsize=40mm\epsffile{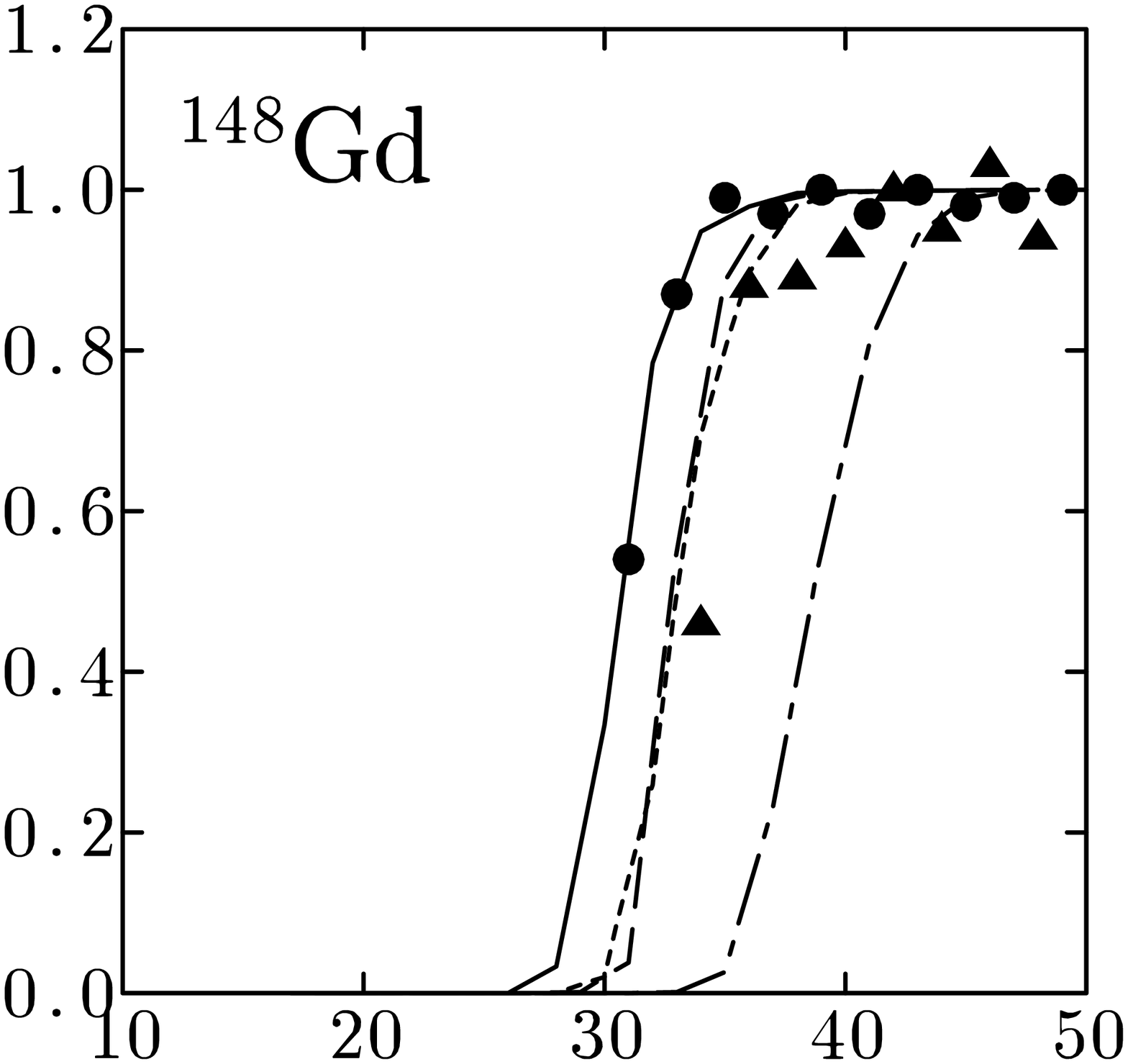} \hspace{-4mm}
\epsfxsize=40mm\epsffile{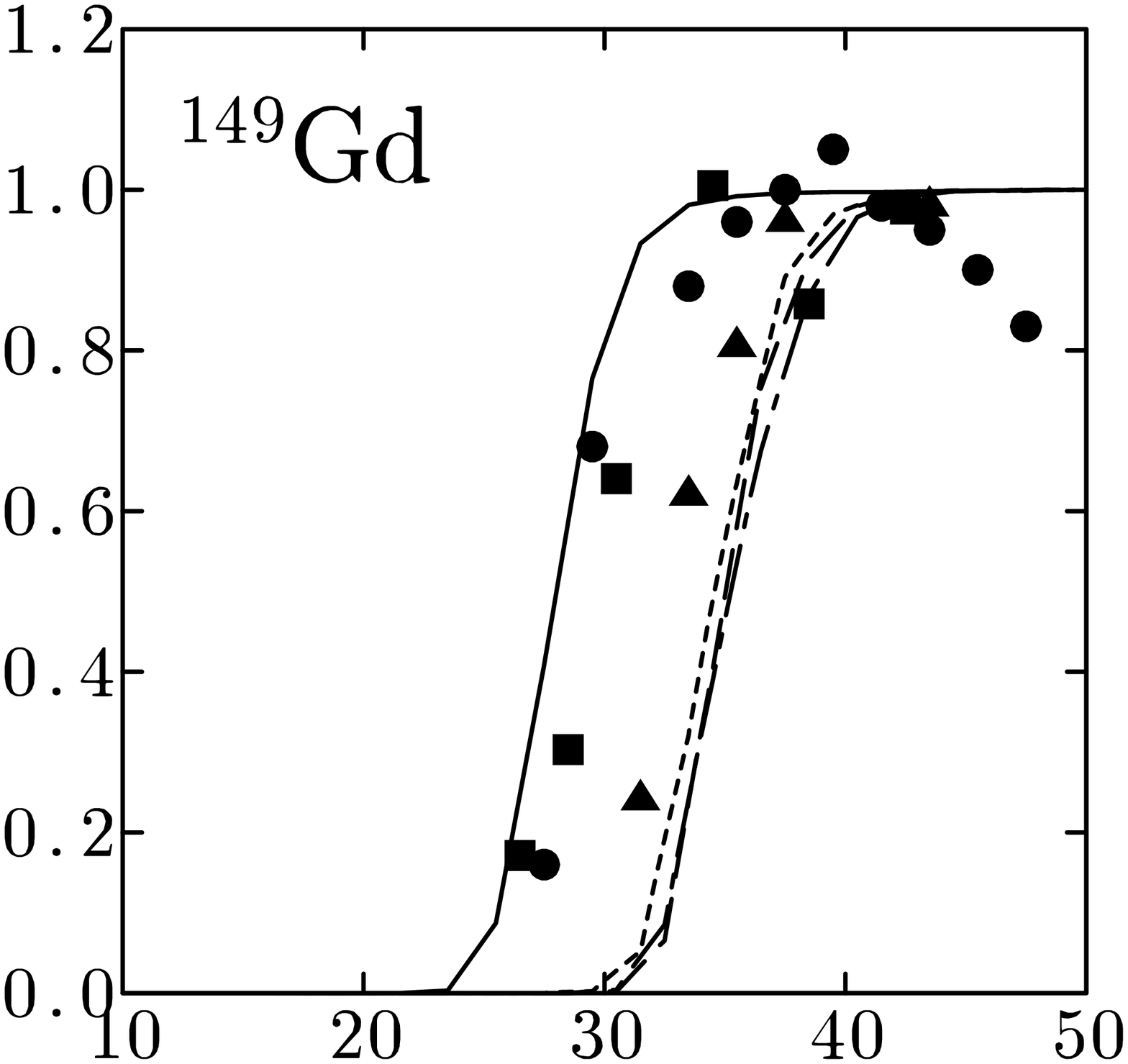} \hspace{-4mm}
\epsfxsize=40mm\epsffile{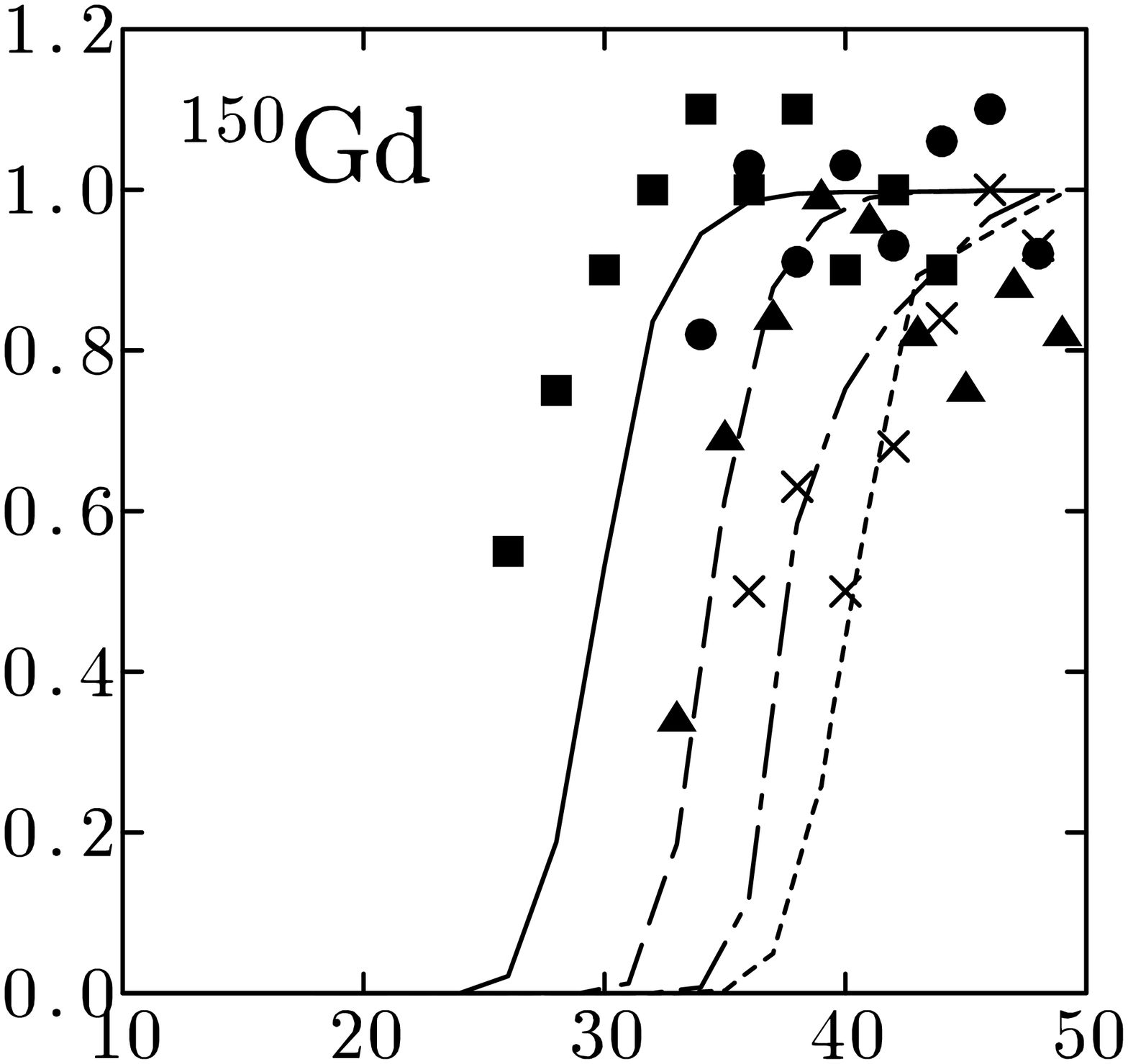} \hspace{-4mm}
}
\vspace{-2mm}
\centerline{
\epsfxsize=40mm\epsffile{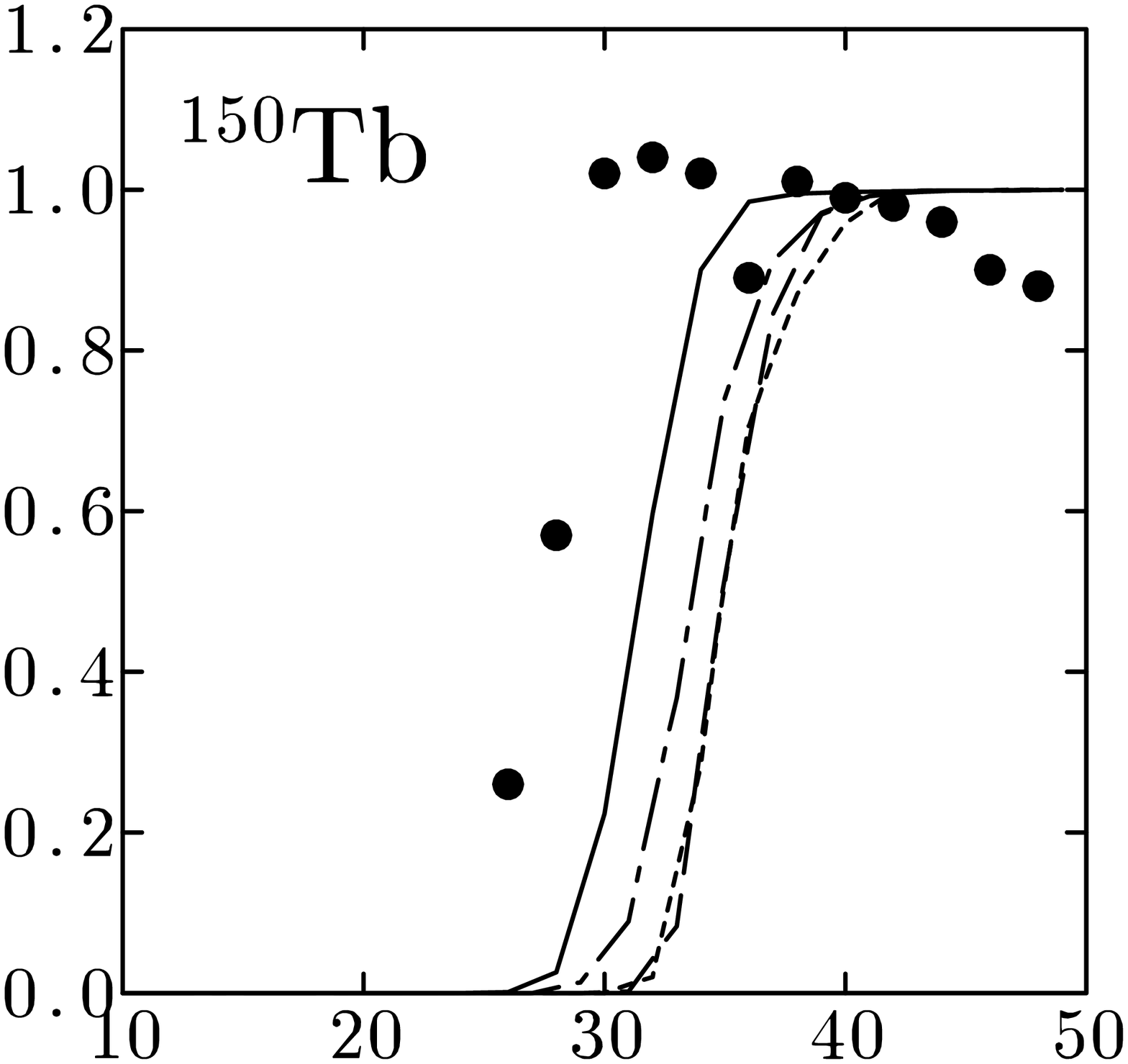} \hspace{-4mm}
\epsfxsize=40mm\epsffile{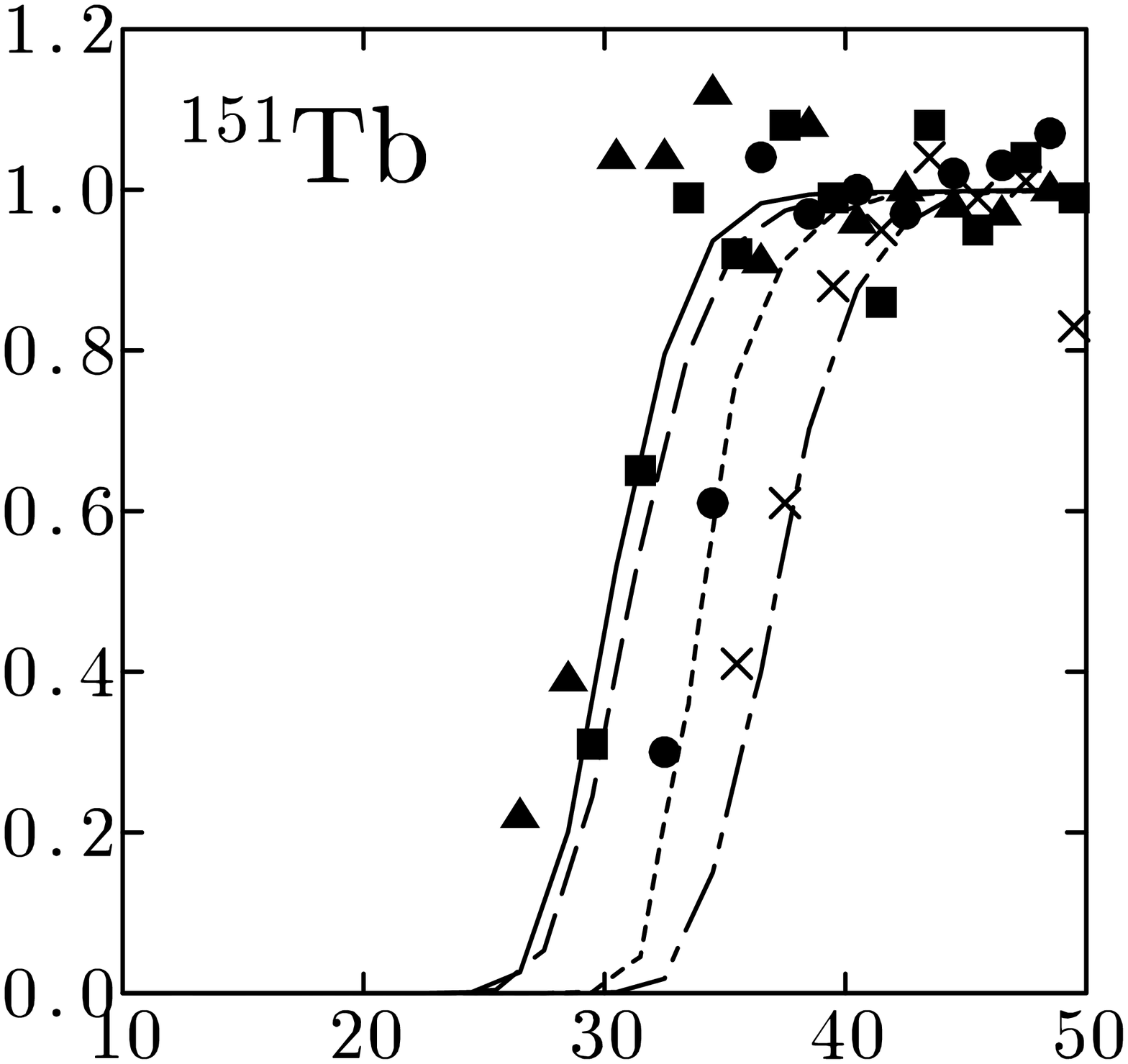} \hspace{-4mm}
\epsfxsize=40mm\epsffile{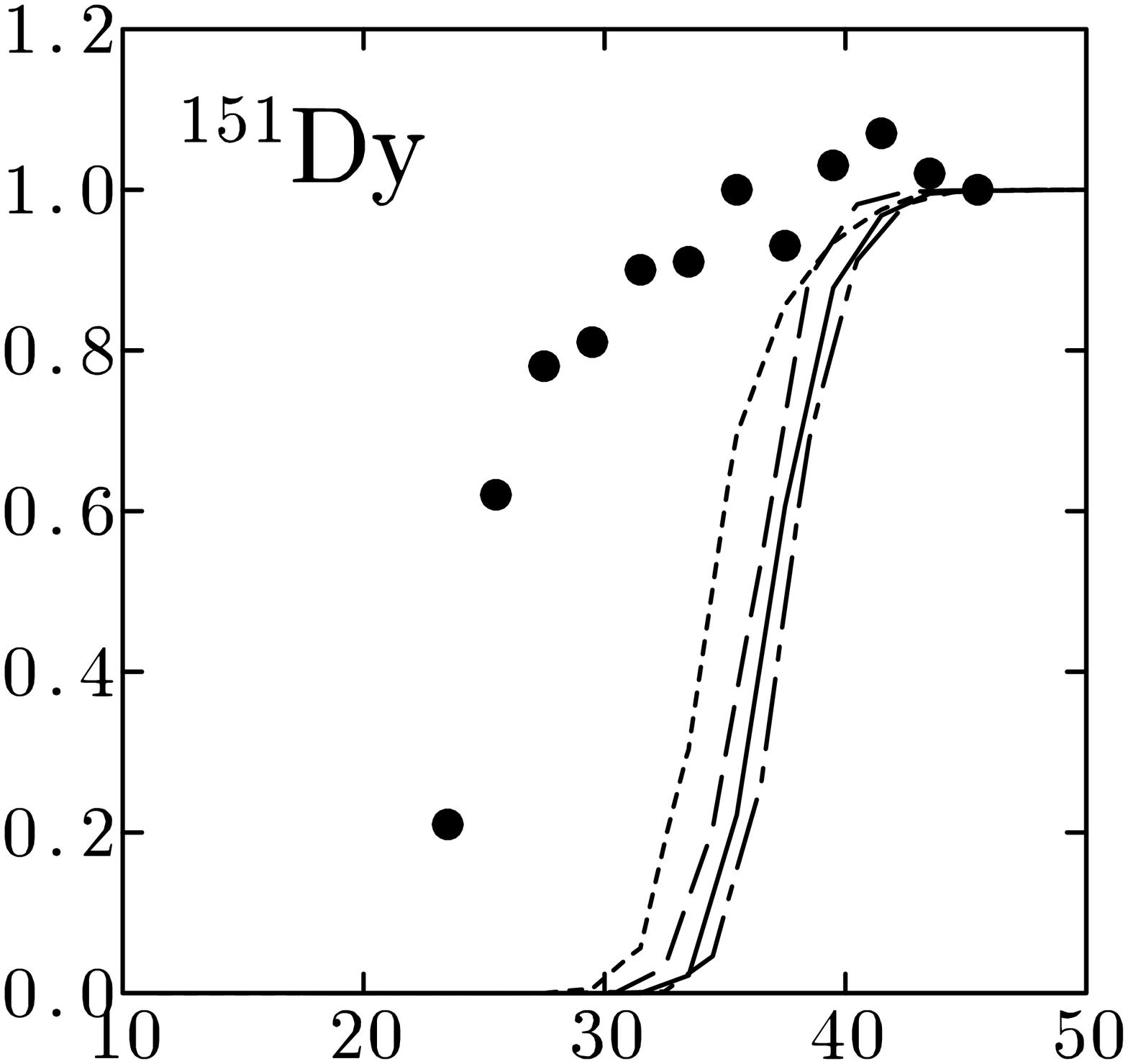} \hspace{-4mm}
\epsfxsize=40mm\epsffile{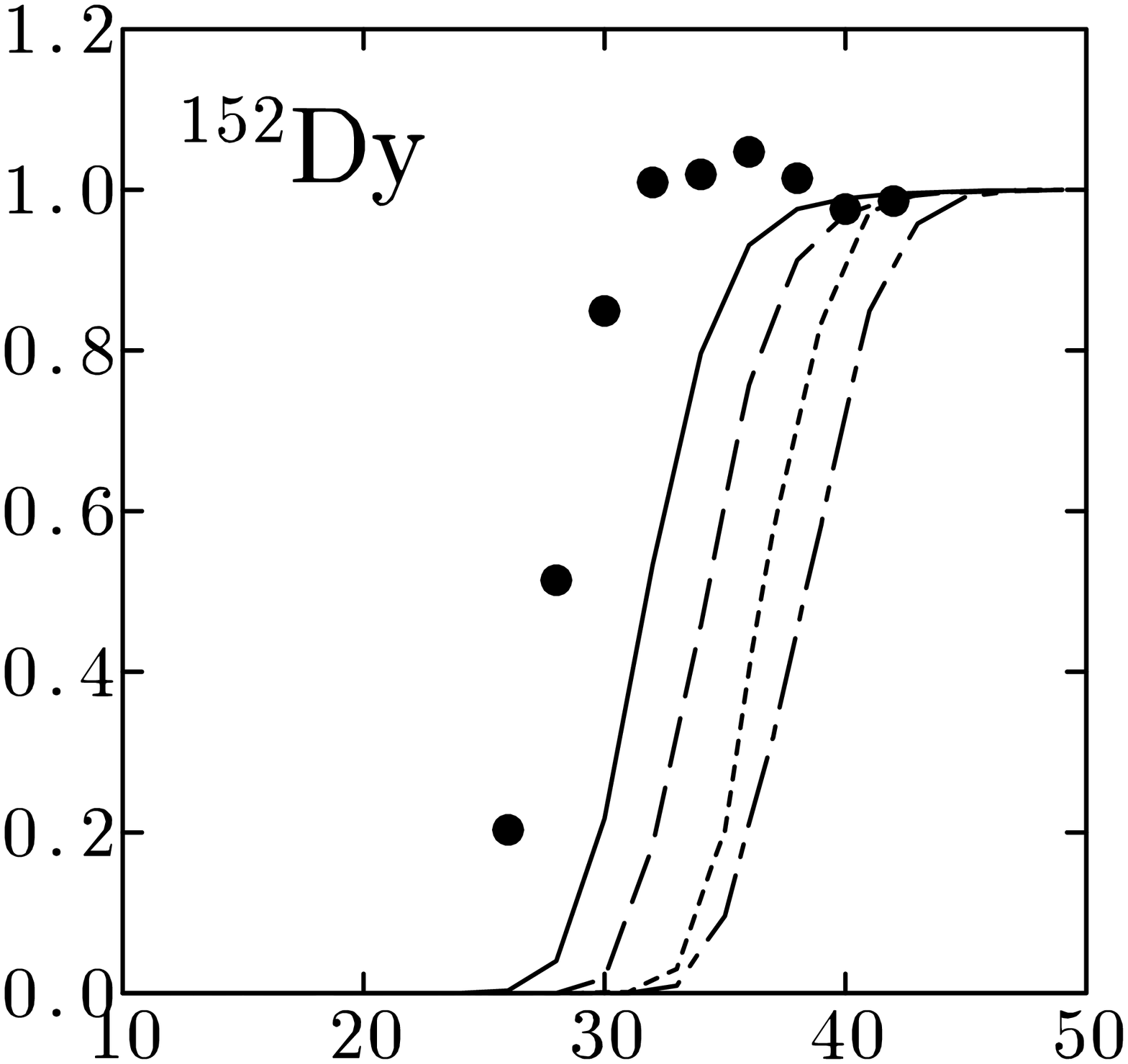} \hspace{-4mm}
}
\vspace{-10mm}
\caption{
 $\gamma$-ray intensities as functions of spin
 for nuclei in the $A \approx 150 $ region.
 Solid, dashed, dotted and dash-dotted curves denote
 calculated ones for bands with different parity and signature
 sorted by excitation energy in the feeding region.
 Circle, triangle, square and cross denote experimental ones
 for yrast and excited bands taken from \protect\cite{SDT97,CFB93}.
 }
\label{fig:A150}
\vspace{-8mm}
\end{figure}

\begin{figure}[htb]
\centerline{
\epsfxsize=40mm\epsffile{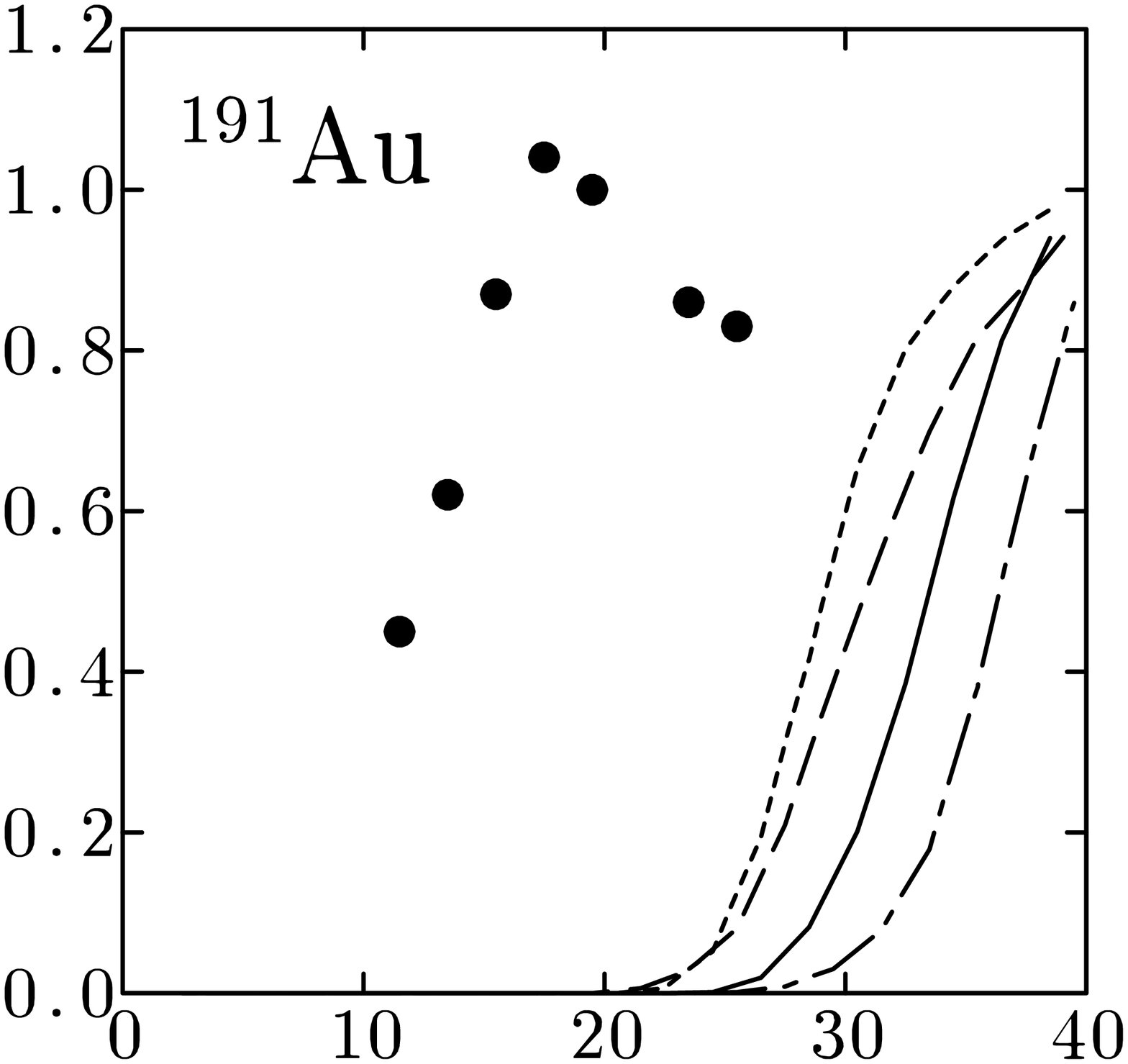} \hspace{-4mm}
\epsfxsize=40mm\epsffile{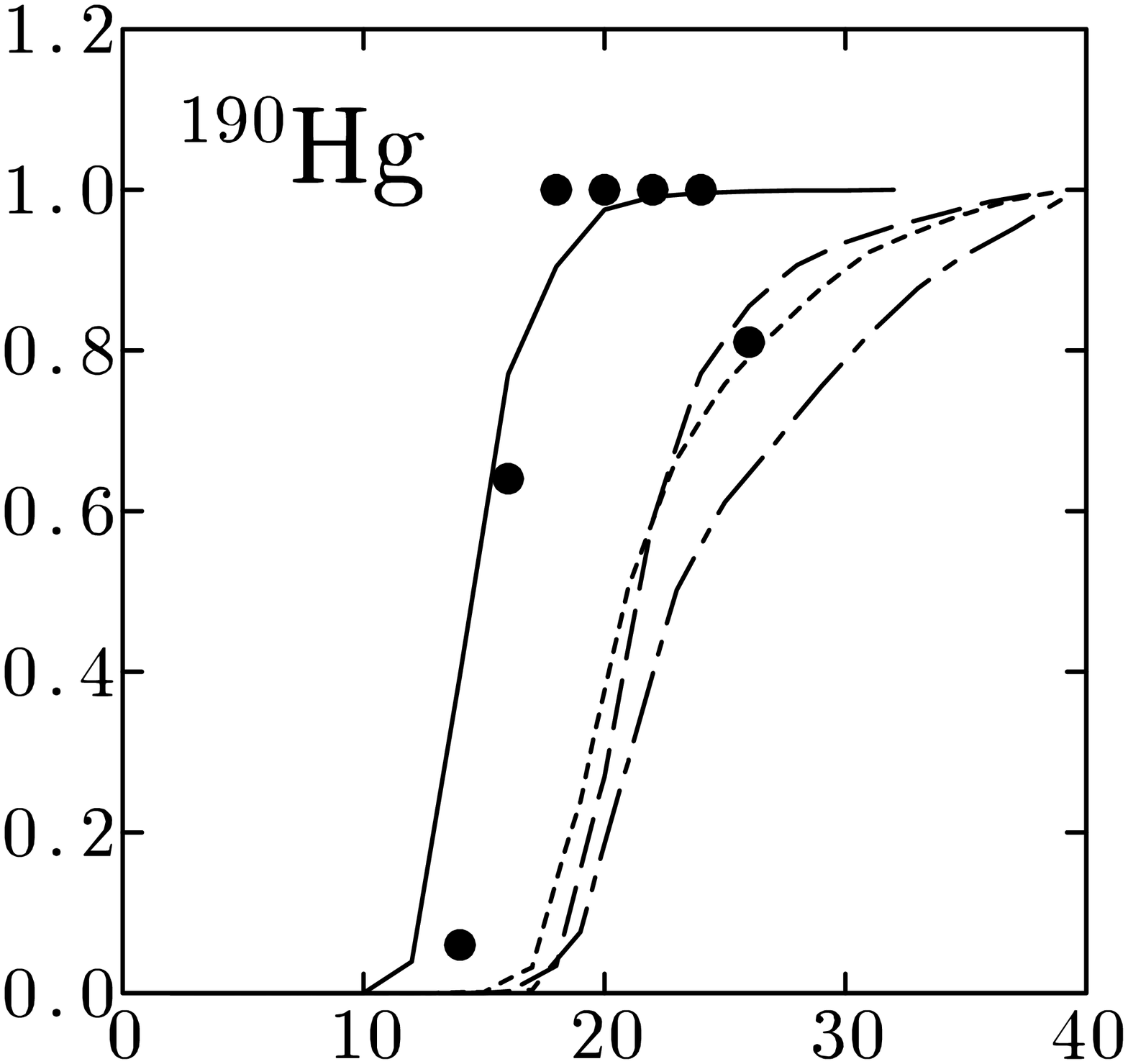} \hspace{-4mm}
\epsfxsize=40mm\epsffile{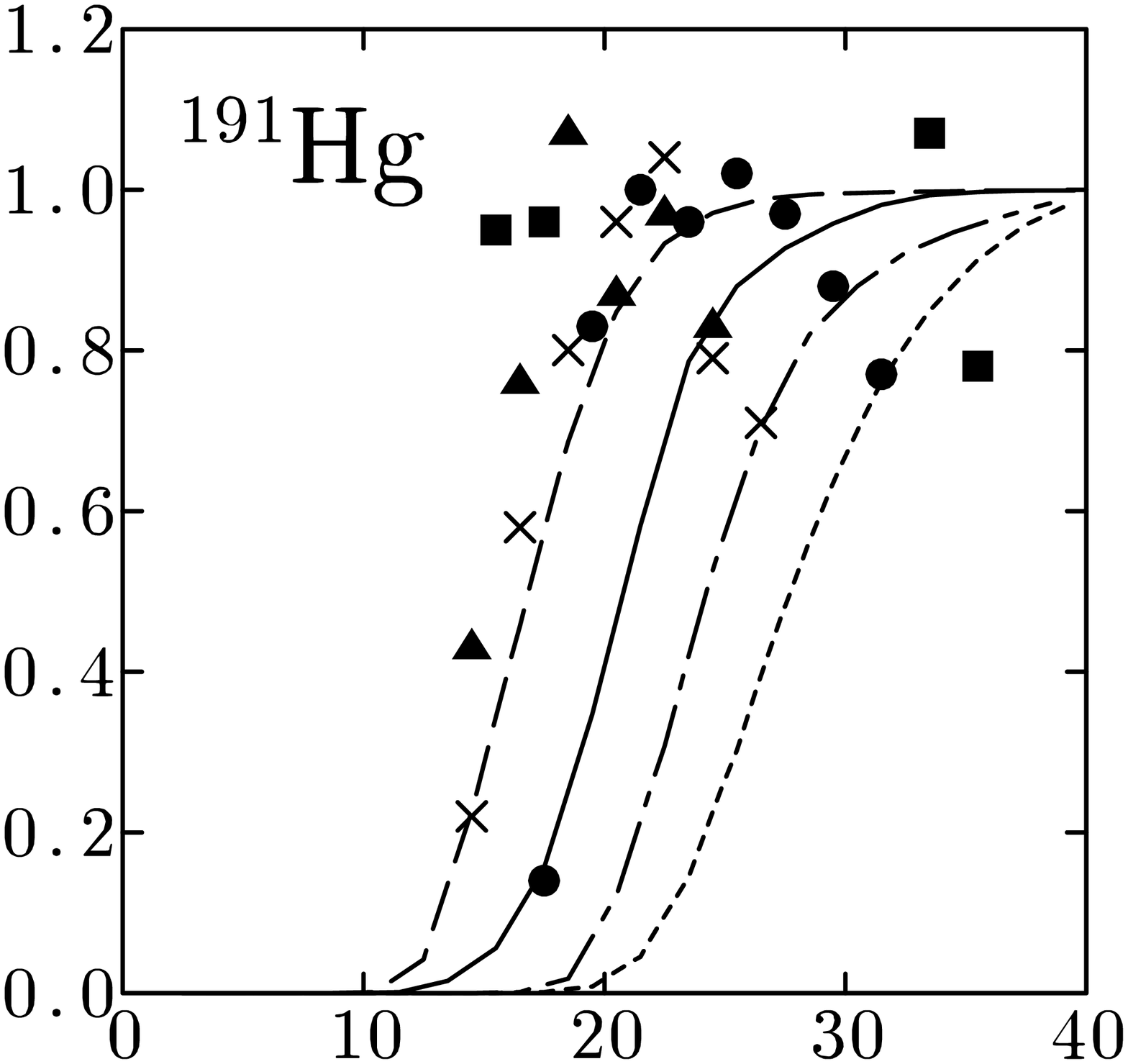} \hspace{-4mm}
\epsfxsize=40mm\epsffile{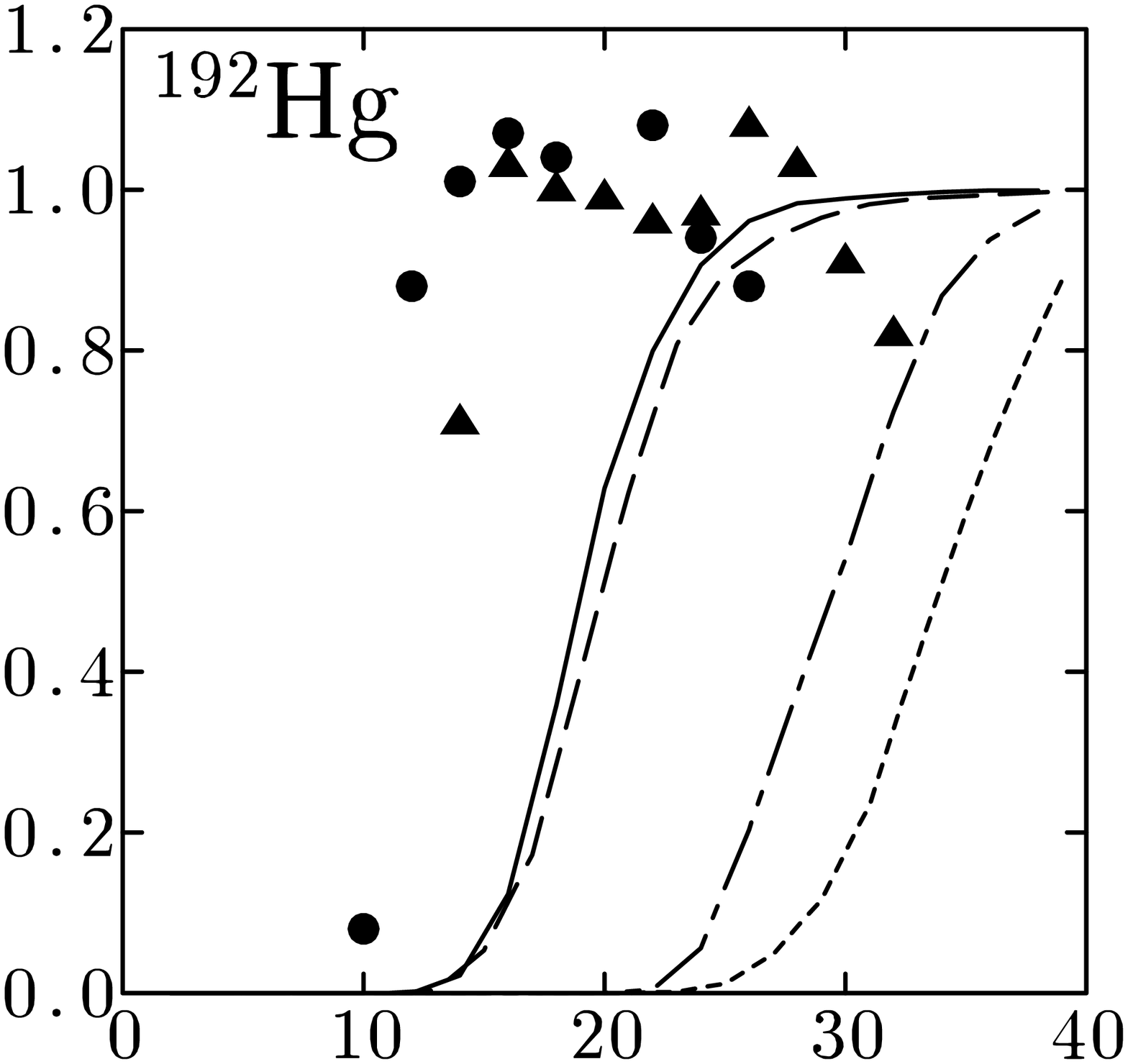} \hspace{-4mm}
}
\vspace{-2mm}
\centerline{
\epsfxsize=40mm\epsffile{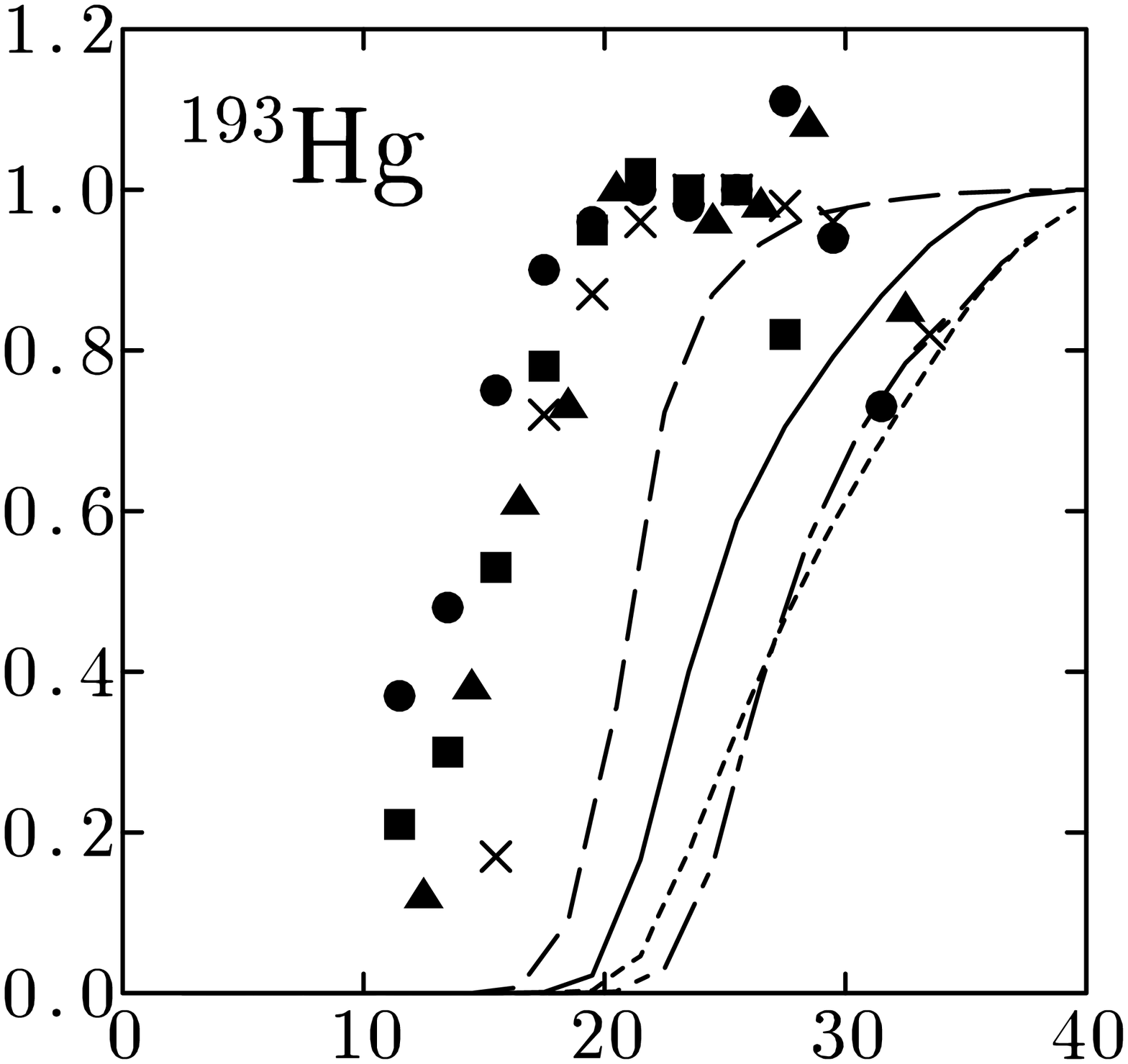} \hspace{-4mm}
\epsfxsize=40mm\epsffile{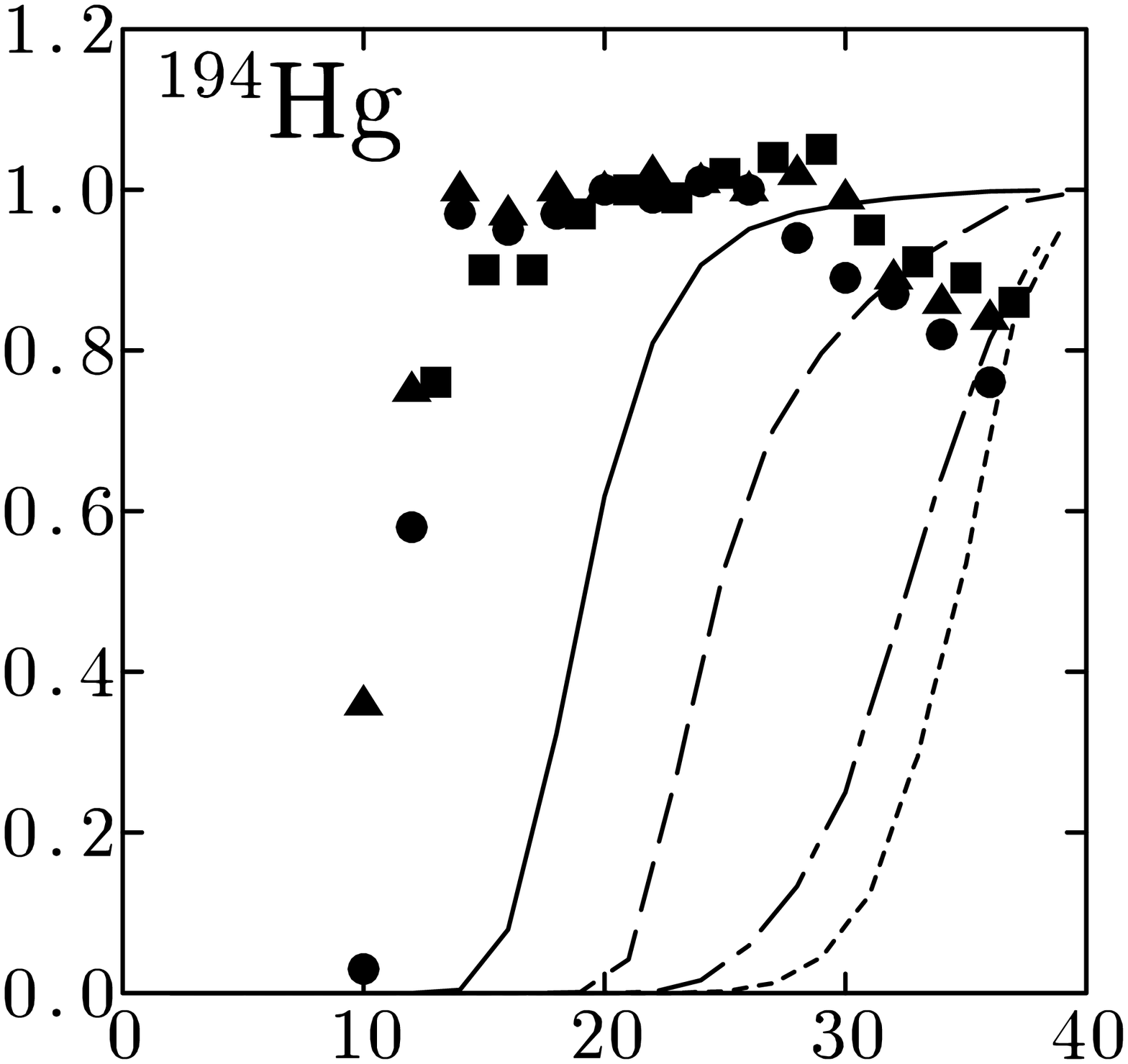} \hspace{-4mm}
\epsfxsize=40mm\epsffile{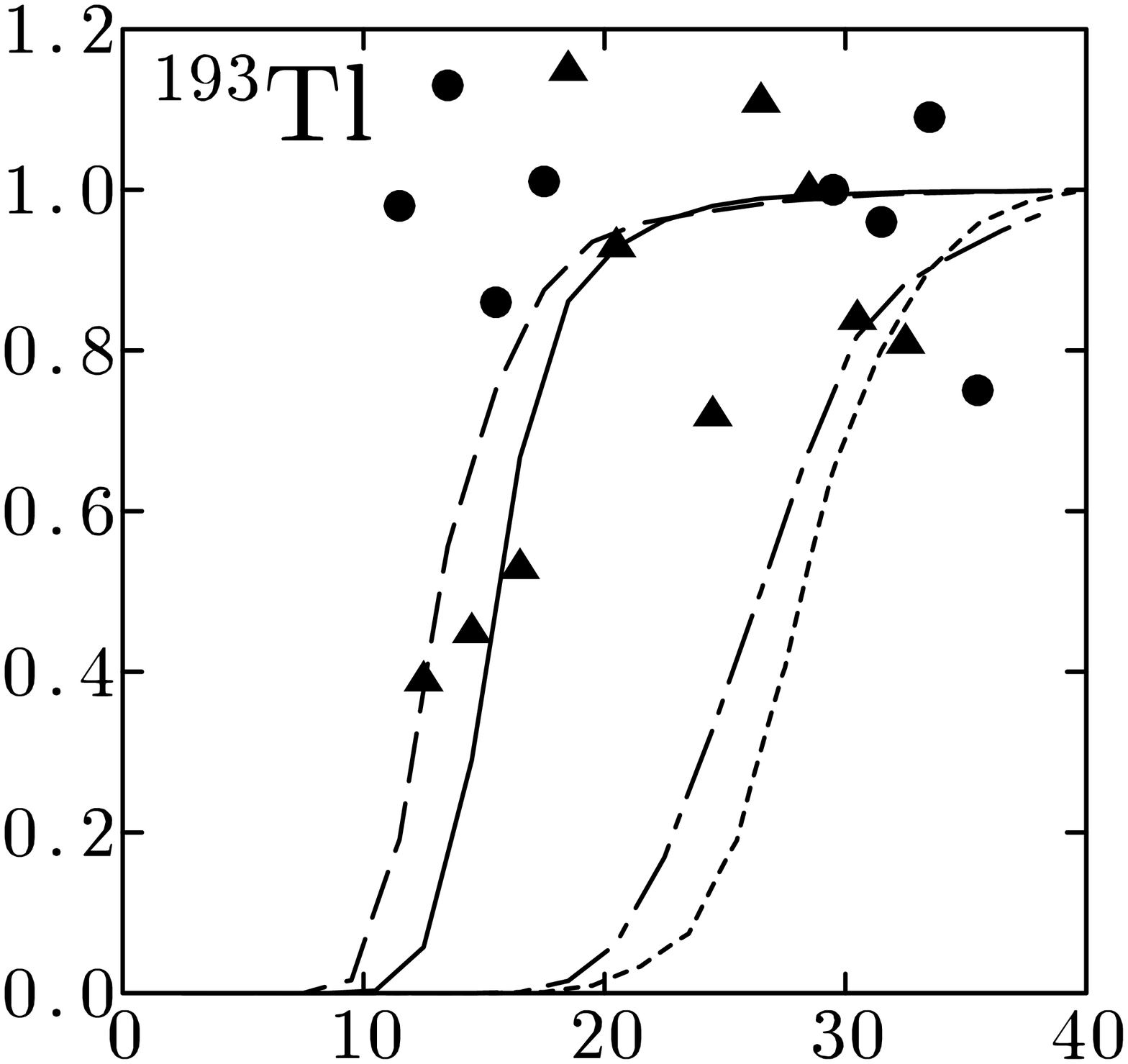} \hspace{-4mm}
\epsfxsize=40mm\epsffile{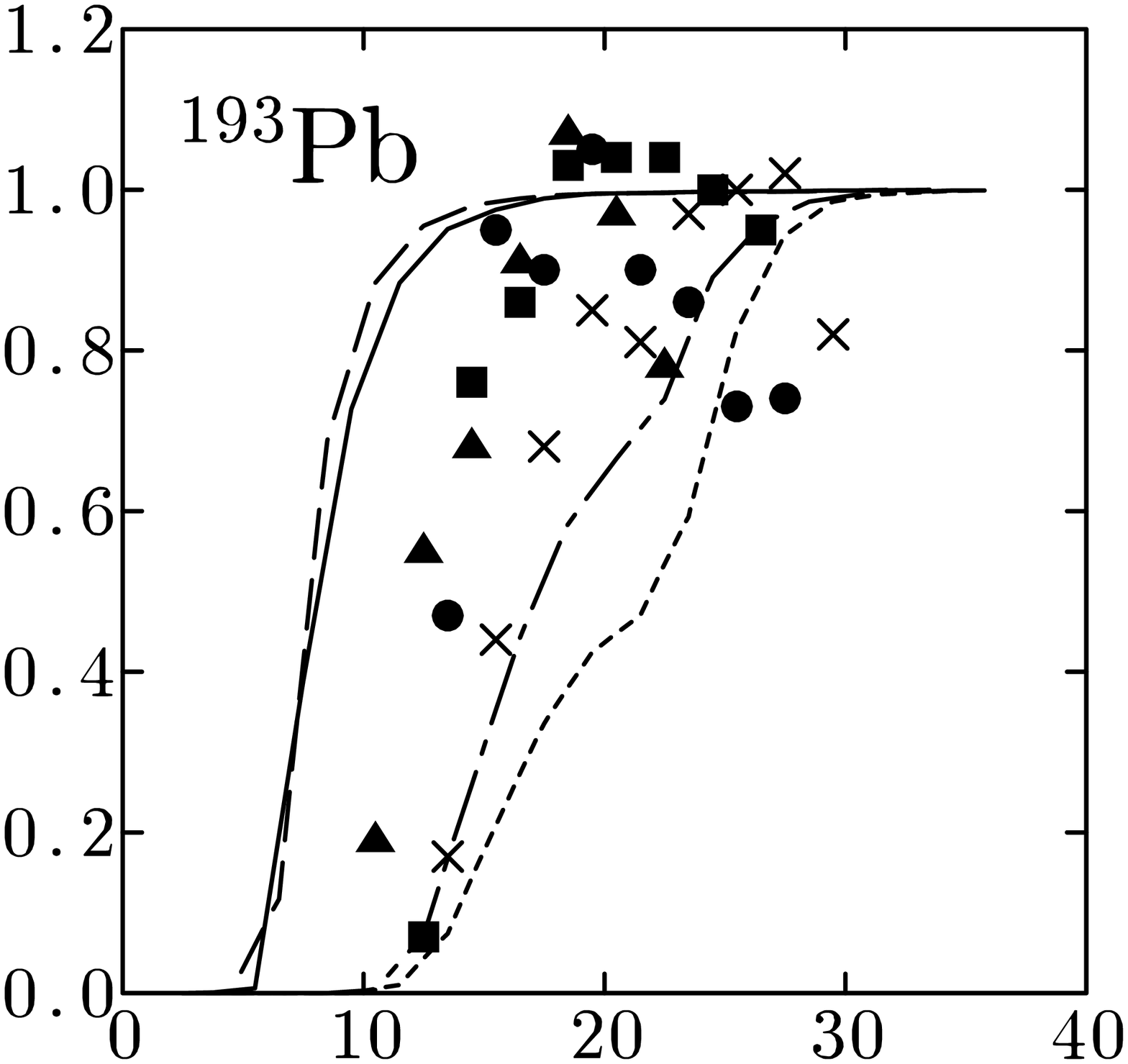} \hspace{-4mm}
}
\vspace{-2mm}
\centerline{
\epsfxsize=40mm\epsffile{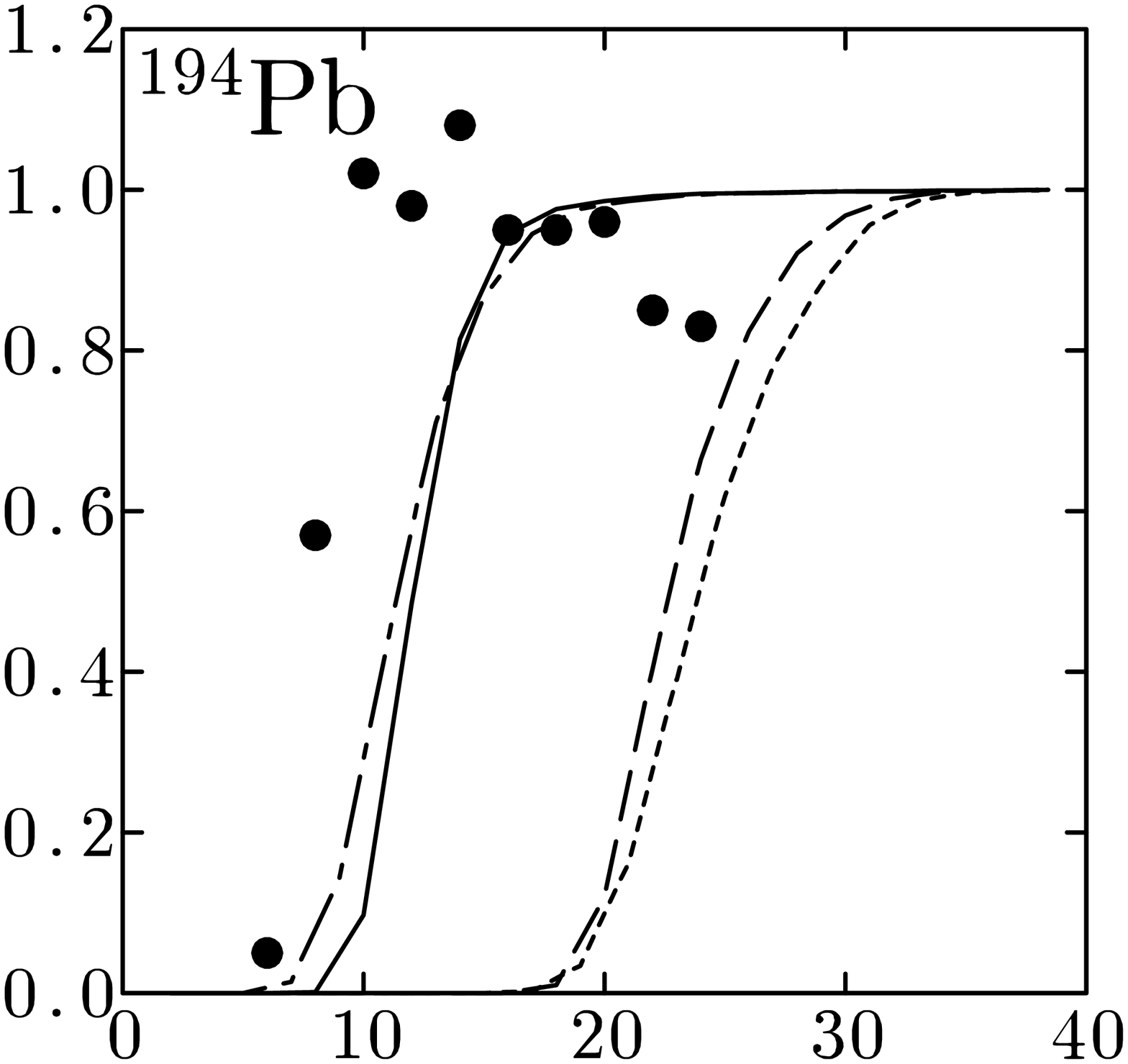} \hspace{-4mm}
\epsfxsize=40mm\epsffile{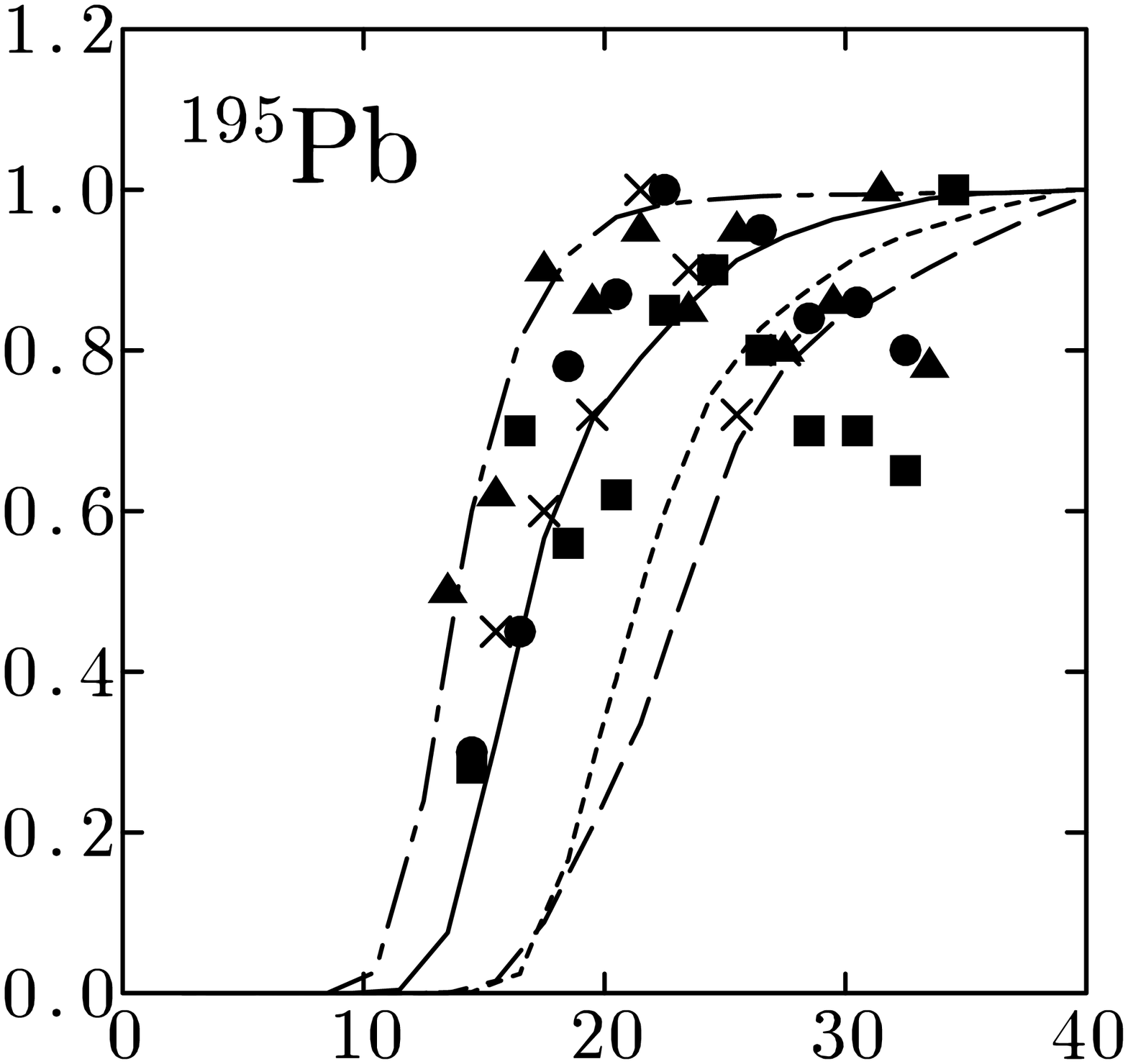} \hspace{-4mm}
\epsfxsize=40mm\epsffile{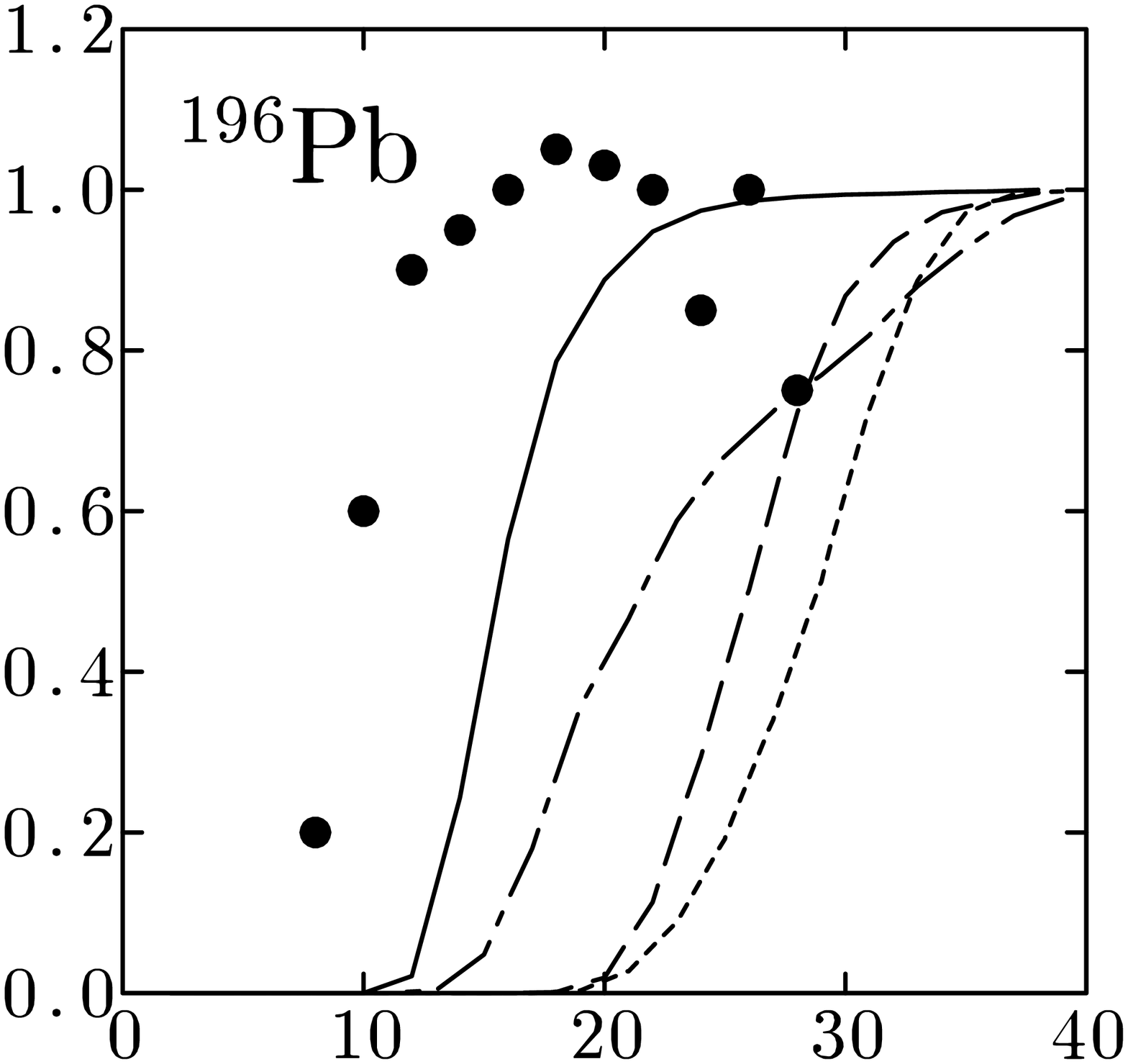} \hspace{-4mm}
\epsfxsize=40mm\epsffile{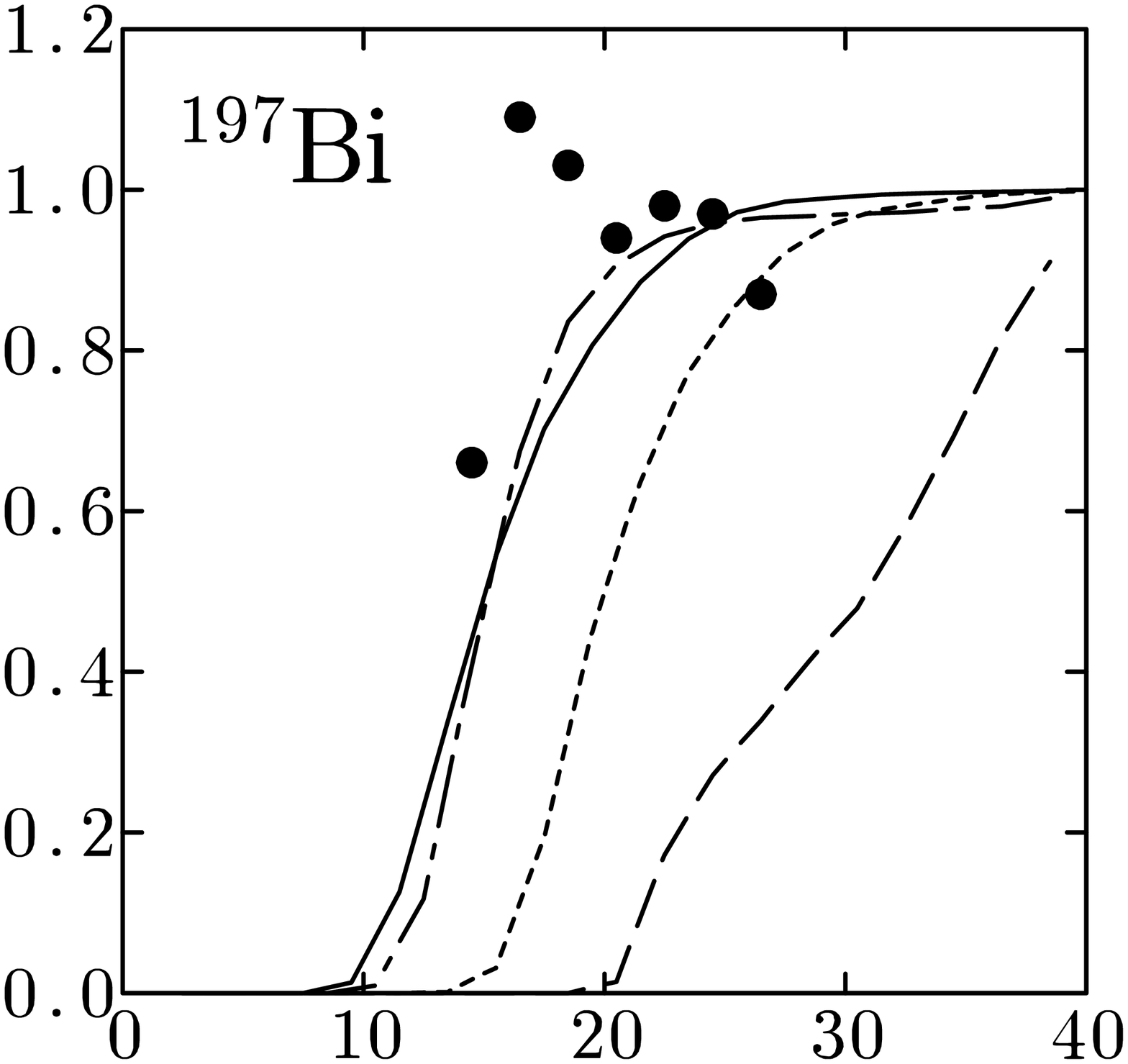} \hspace{-4mm}
}
\vspace{-10mm}
\caption{
 Same as Fig.~\protect\ref{fig:A150} but for nuclei
 in the $A \approx 190 $ region.
 Here the back-shift of 1 MeV is employed throughout
 in the level density formula.
 }
\label{fig:A190}
\vspace{-7mm}
\end{figure}

Decay of superdeformed (SD) rotational bands
out to normal deformed (ND) states is one of
the most interesting nuclear structure problems.
It can be viewed as a shape-coexistence phenomena,
and the mixing mechanism of two states having very different
internal structures can be studied as functions of angular momenta
and excitation energies.
Using the compound mixing model~\cite{VBD90},
we have investigated the decay-out phenomena in~\cite{SDVB93},
where a consistent description was presented and the rapid
decay-out was well understood.
After the calculation of~\cite{SDVB93},
more systematic and improved calculations have been performed~\cite{SDV96},
in which all the relevant quantities to the decay-out have been
calculated without any adjustable parameters.
We have found, however, it is rather difficult to reproduce
the decay-out spin for individual SD bands
(cf. Figs.~\ref{fig:A150},~\ref{fig:A190}).

The basic idea of our framework~\cite{VBD90,SDVB93} is based
on a simple two-well-mixing:  The two groups of unperturbed states
belonging to wells, I (e.g. ND) and II (e.g. SD),
are separated by a potential barrier in some collective
coordinate space (e.g. deformations).
The most important is the (mean) coupling strength between the two,
and is estimated as~\cite{BL80}
\begin{displaymath}
  v^2_\IItoI = \left\{
   \begin{array}{ll}
    (\hbar\omega_\I/2\pi)(\hbar\omega_\II/2\pi)\, T_\IItoI,
     & (\mbox{regular-regular}), \hspace{56.5mm} (\mbox{1a})\\
    (D_\I/2\pi)(\hbar\omega_\II/2\pi)\, T_\IItoI,
     & (\mbox{chaos-regular}), \hspace{59.1mm} (\mbox{1b})\\
    (D_\I/2\pi)(D_\II/2\pi)\, T_\IItoI ,
     & (\mbox{chaos-chaos}),\hspace{62.1mm} (\mbox{1c}) \\
   \end{array} \right.
\label{eq:vcoupl}
\end{displaymath}
\addtocounter{equation}{1}
where $\omega_{\I,\II}$ is the frequency of each well,
$D_{\I,\II}=1/\rho_{\I,\II}$ is the mean level distance,
and $T_\IItoI=(1+\exp{2S_\IItoI})^{-1}$ is the transmission coefficient
of the tunneling process between the wells with the least action $S_\IItoI$.
Depending on the situation of states in each well, one of three cases
should be chosen in Eq.~(\ref{eq:vcoupl}):
An example of~(1a) is the decay of high-$K$ isomers~\cite{NSS96},
in which severe breakdown of $K$-hindrance has been observed recently.
The present problem of decay-out of near-yrast SD bands
corresponds to~(1b), where the ND states are compound states
and described by the GOE model~\cite{VBD90}.
In higher energies thermally excited SD bands
are rotationally damped~\cite{YM98},
and (1c) may be most appropriate for
the decay-out of such SD continuum states~\cite{YMS00}.

In the present case,
the relative intensity of $\gamma$-rays inside the SD band,
which is calculated by the total decay-out probability $\Nout(I)$ at spin $I$,
is almost the only observable and will be discussed in the followings.
According to~\cite{VBD90}, $\Nout$ is determined by
combinations, $\Gammat/D_\rmn$ and $\Gammas/\Gamman$,
of four quantities; the spreading width $\Gammat$=$2\pi v^2_\ston/D_\rmn$
due to the coupling~(\ref{eq:vcoupl}), the level density $1/D_\rmn$
of the ND compound states, and the $\gamma$-decay widths
$\Gammas$ and $\Gamman$ of the SD and ND bands, respectively,
where $\Gammas$ is of rotational E2,
while $\Gamman$ is mainly of statistical E1
(see~\cite{SDVB93,YMS00} for details).
It is worth mentioning that the model of~\cite{VBD90}
for $\Nout$ was re-examined by means of
a statistical model of compound nucleus~\cite{GW99},
and both models were found to give identical results for
actual range of four parameters being relevant
to decay of both the $A \approx 150$ and 190 SD nuclei.

In this talk, we present the result of systematic calculations for
the relative $\gamma$-ray intensity as a function of
angular momentum in comparison with experimental data.
We also discuss how the decay-out spin is
understood from the calculations.
Possible directions for improvements of the theoretical framework
are suggested further.

\section{RESULTS AND DISCUSSIONS}

Quantities to be calculated are $\Gammat$, $D_\rmn$,
$\Gammas$, and $\Gamman$ as functions of spin.
$\Gammas$ is the usual rotational E2 width,
while the level density
and the statistical E1 width are given by
\begin{equation}
   \rho_\FG (U) = (\sqrt{\pi}/48)\,a^{-\frac{1}{4}}
      U^{-\frac{5}{4}}\exp{2\sqrt{aU}}, \quad
 \mGamma^\Eone_\rmn=0.15 \times 2.3 \times 10^{-11} \, NZ A^{1/3}\,
       (U/a)^{5/2},
\label{eq:RhoEone}
\end{equation}
in unit of MeV, where $U$ is excitation energy of the SD band
from the ND yrast state, and $a$ parameter is taken from
an empirical analysis of~\cite{ExpLvlD}
including the shell and temperature effects.
The decay-out spin $\Iout$ in the $A \approx 190$ region is generally low
so that the back-shift is used for $A \approx 190$ nuclei,
i.e. $U$ is replaced by $U-1$ MeV in $\rho_\FG$,
throughout the present investigation.
As for $\Gammat$, the least action path in the $(\epsilon_2,\gamma)$-plane
is solved and the frequency $\omegas$
and the action $S_\ston$ along it are calculated.
Therefore all what we need are excitation energies of both SD and ND
rotational bands, their quadrupole moments, and the potential energy
surface and the mass tensor.  We have used the Nilsson-Strutinsky
calculation for potential
with the pairing correlations included in the RPA order,
and the pairing hopping model for mass tensor~\cite{BBBV90};
see~\cite{YMS00} for the detailed formulation.

In Figs.~\ref{fig:A150} and~\ref{fig:A190}
the results of calculation for relative intensities are shown,
where those for the lowest band in each parity and signature
(four bands) are included in one nucleus.
The basic characteristics of intensity pattern are
reproduced in both the $A \approx 150$ and 190 regions;
especially the rapid decrease of transitions at lower spins.
Apparently, however, the decay-out spin does not agree precisely,
and the detailed features,
like the relative ordering of decay-out for excited bands in one nucleus,
or relative difference of neighbouring nuclei, are not well described.
Comparing Figs.~\ref{fig:A150} and~\ref{fig:A190},
decay-out spins are rather well reproduced
on average for the $A \approx 150$ nuclei,
while the calculated $\Iout$'s are still higher than the observed ones
for the $A \approx 190$ nuclei,
even though the back-shift is used for them.
Although there are some cases
where the yrast SD band decays at higher spin than the excited ones
in Figs.~\ref{fig:A150} and~\ref{fig:A190},
it is because the energy ordering is determined in the feeding-spin region;
the `real' yrast band survives longest in most cases.

\begin{figure}[t]
\centerline{
\epsfxsize=40mm\epsffile{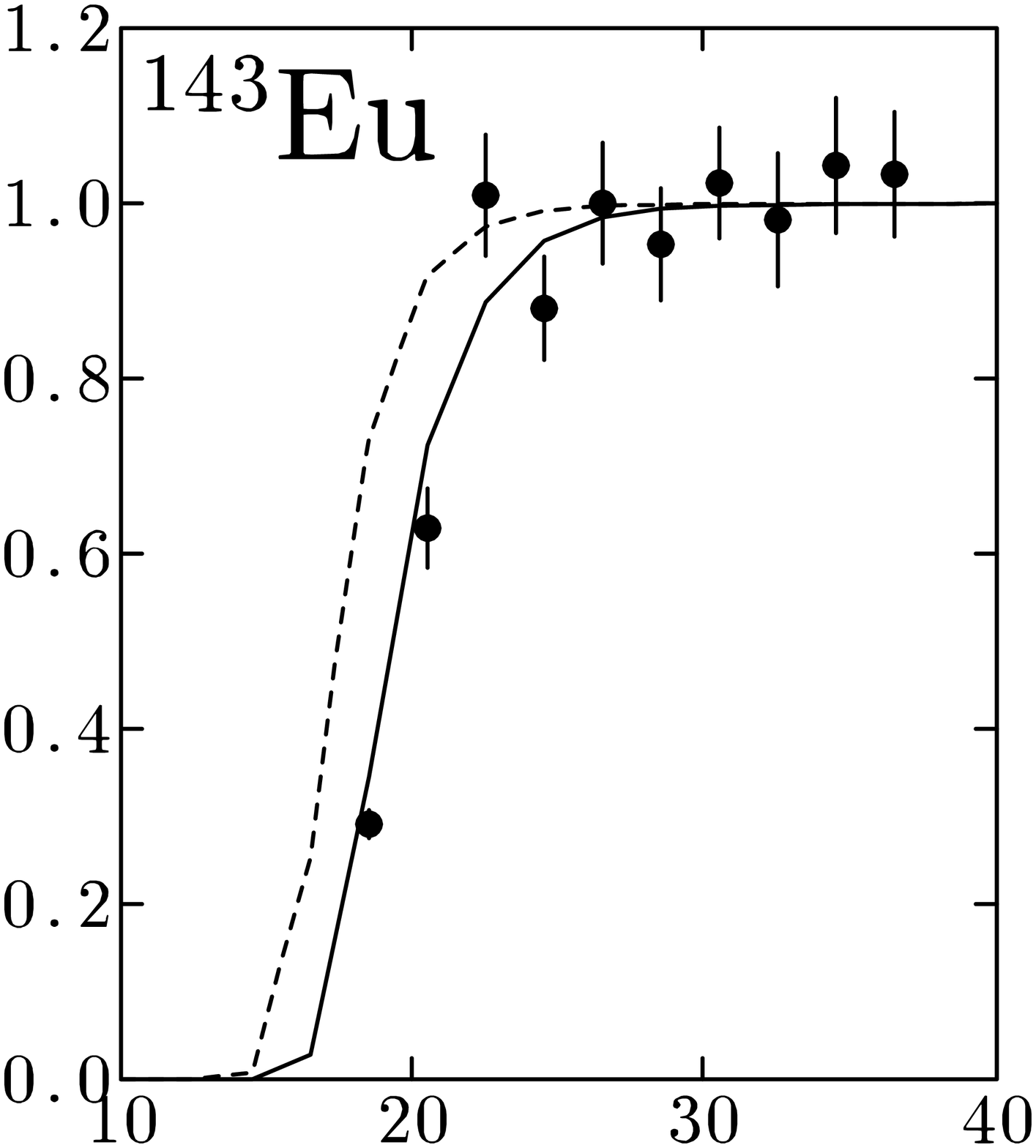} \hspace{-4mm}
\epsfxsize=40mm\epsffile{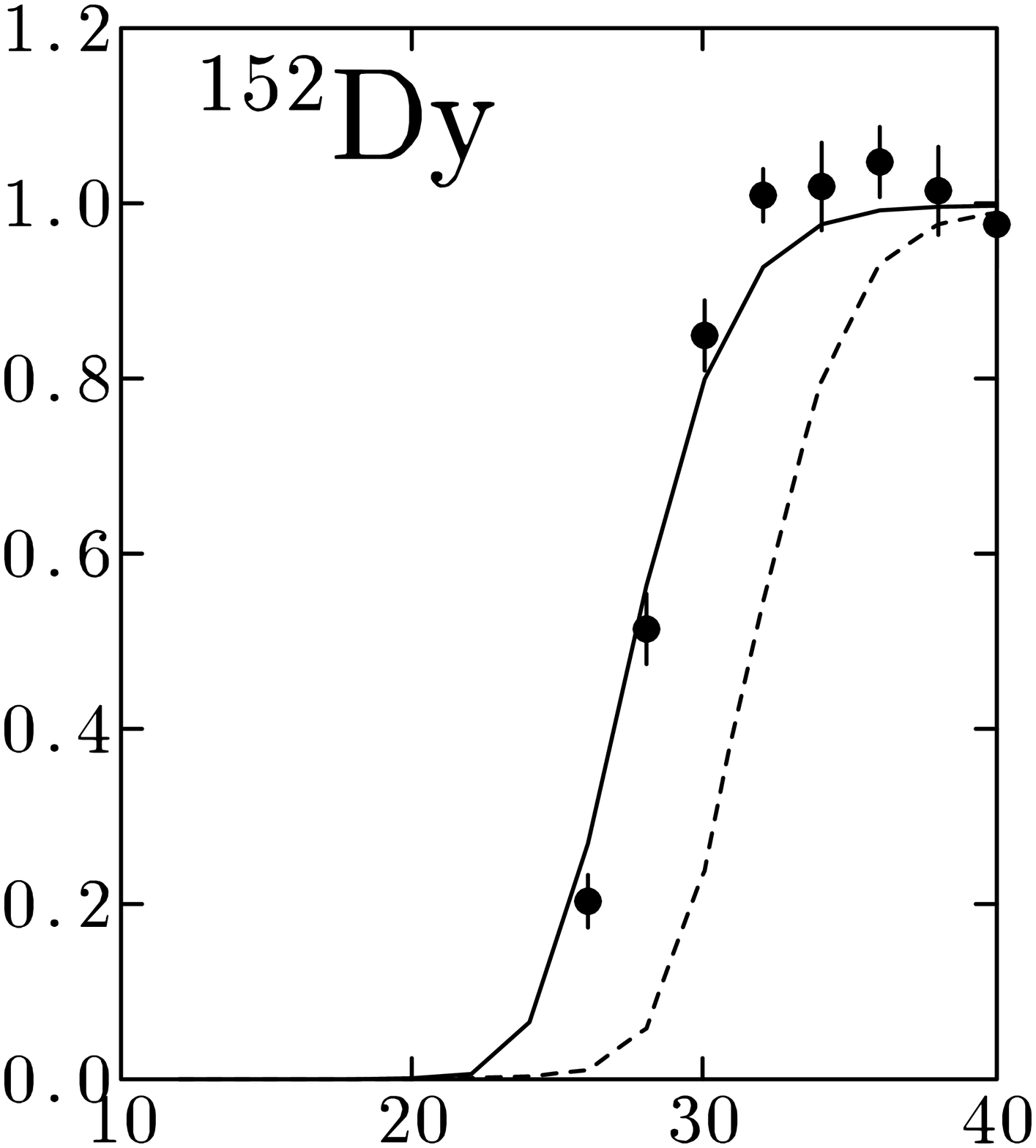} \hspace{-4mm}
\epsfxsize=40mm\epsffile{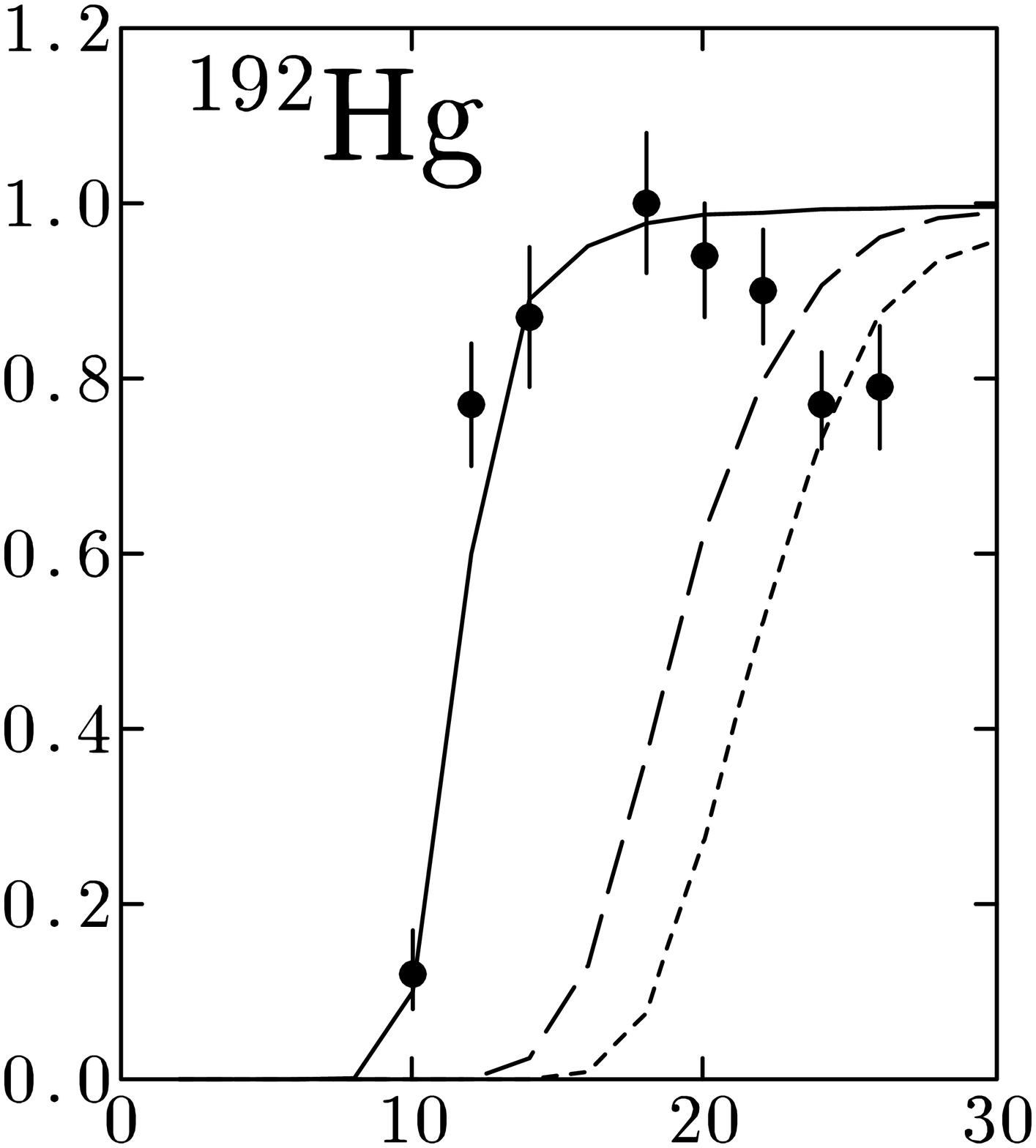} \hspace{-4mm}
\epsfxsize=40mm\epsffile{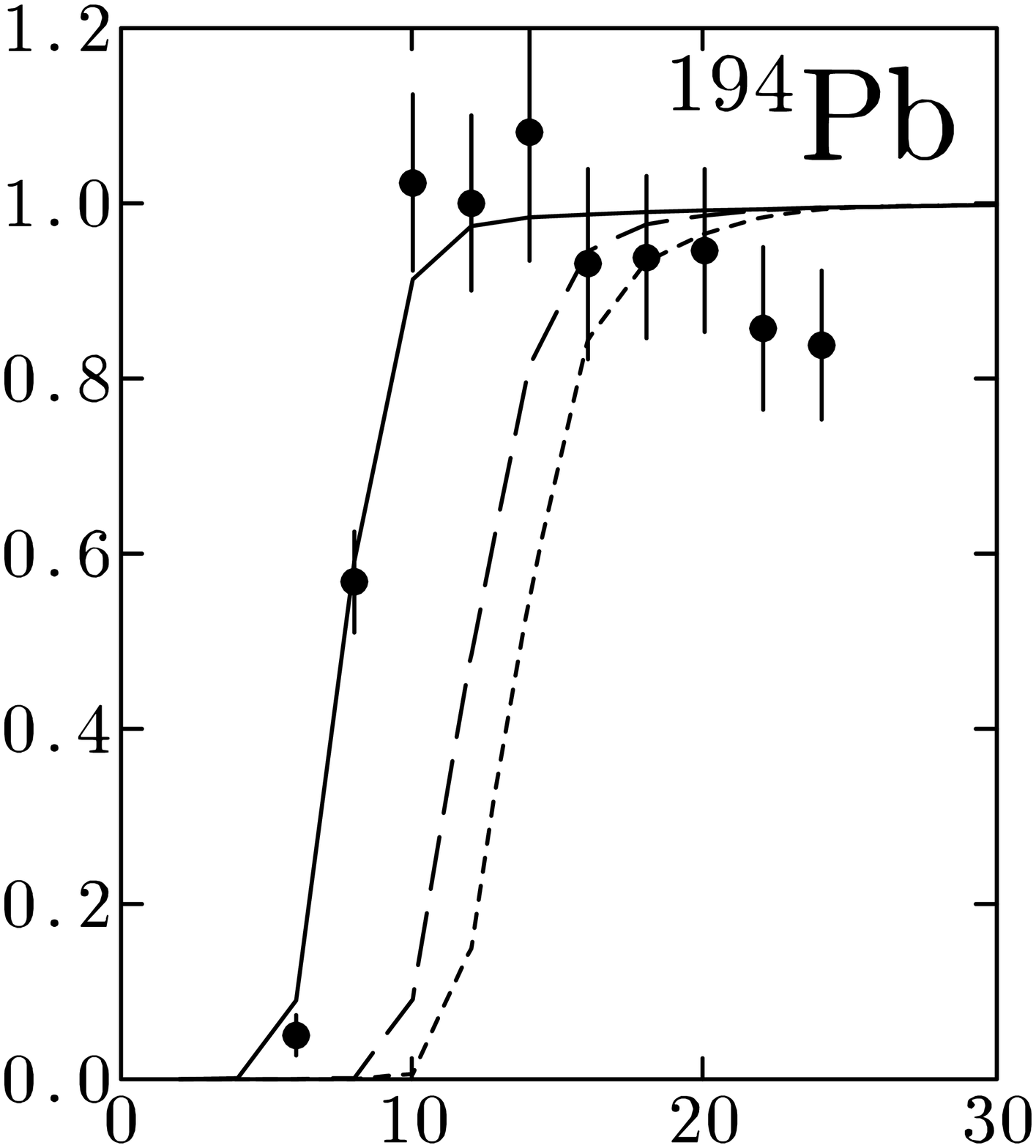} \hspace{-4mm}
}
\vspace{-10mm}
\caption{
 Results with fitting the decay-out spin by scaling
 the mass tensor (see Table~\protect\ref{tab:Gamma}).
 Dotted (dashed) curves denote results with no adjustments
 (1 MeV back-shift).}
\label{fig:FitMass}
\vspace{-8mm}
\end{figure}

As is shown in Figs.~\ref{fig:A150} and~\ref{fig:A190},
the decay-out spin is not precisely reproduced in individual cases:
What are the reasons?   Possible sources of ambiguity in our model
may be the level density and the mass tensor;
we believe that the potential energy surface by the Nilsson-Strutinsky
method is reliable.
As for the change of the level density~\cite{SDVB93} it was found that,
for example, a factor about $10^{-1}$ is enough to fit $\Iout$ in $^{152}$Dy,
while factor $10^{-3}-10^{-4}$ is required for $^{192}$Hg;
the latter may be out of the range of allowed ambiguity.
The mass tensor is calculated by using the Fermi-gas estimate
of the number of level-crossings~\cite{BBBV90},
and detailed properties of individual nucleus is not included.
Therefore, we try to use a mass tensor multiplied by
a scaling factor $\Cmass$ and adjust it so as to fit $\Iout$.
Resultant intensities for four selected nuclei
are shown in Fig.~\ref{fig:FitMass},
and the values of $\Cmass$ and of various quantities at $\Iout$
are summarized in Table~\ref{tab:Gamma}.
Again, the correction factor $\Cmass$ for $A \approx 150$ nucleus
is rather small but is considerably larger for $A \approx 190$ nucleus.

\begin{table}[t]
\caption{
 Calculated quantities at the decay-out spin which is fitted
 by scaling the mass (Fig.~\ref{fig:FitMass}).
}
\label{tab:Gamma}
\newcommand{\m}{\hphantom{$-$}}
\newcommand{\cc}[1]{\multicolumn{1}{c}{#1}}
\begin{center}
\setlength{\tabcolsep}{5.3pt}
\begin{tabular}{ccccccccccc}
\hline
   & $\Cmass$ & $\Iout$
   & \cc{$\Gammat$(eV)} & \cc{$\Dn$(eV)}
   & \cc{$\Gammas$(meV)} & \cc{$\Gamman$(meV)}
   & \cc{$S_\ston(\hbar)$} & \cc{$U$(MeV)}  & \cc{$\Nout$} \\
\hline
$^{143}$Eu & 0.7 & $37/2$  & 390  & 630  & 1.5  & 1.3 & 3.3 & 2.57 & .52 \\
$^{152}$Dy & 1.6 & $28$    & 2.2  & 4.8  & 18   & 4.7 & 5.6 & 4.68 & .29 \\
$^{192}$Hg & 2.3 & $12$    & .11  & 69   & .058 & 5.2 & 6.9 & 4.24 & .33 \\
$^{194}$Pb & 2.0 & $ 8$    & .061 & 134  & .011 & 4.3 & 7.2 & 3.89 & .35 \\
\hline
\end{tabular}
\end{center}
\vspace{-12mm}
\end{table}

\begin{figure}[t]
\begin{minipage}[t]{76mm}
\centerline{
\epsfxsize=70mm\epsffile{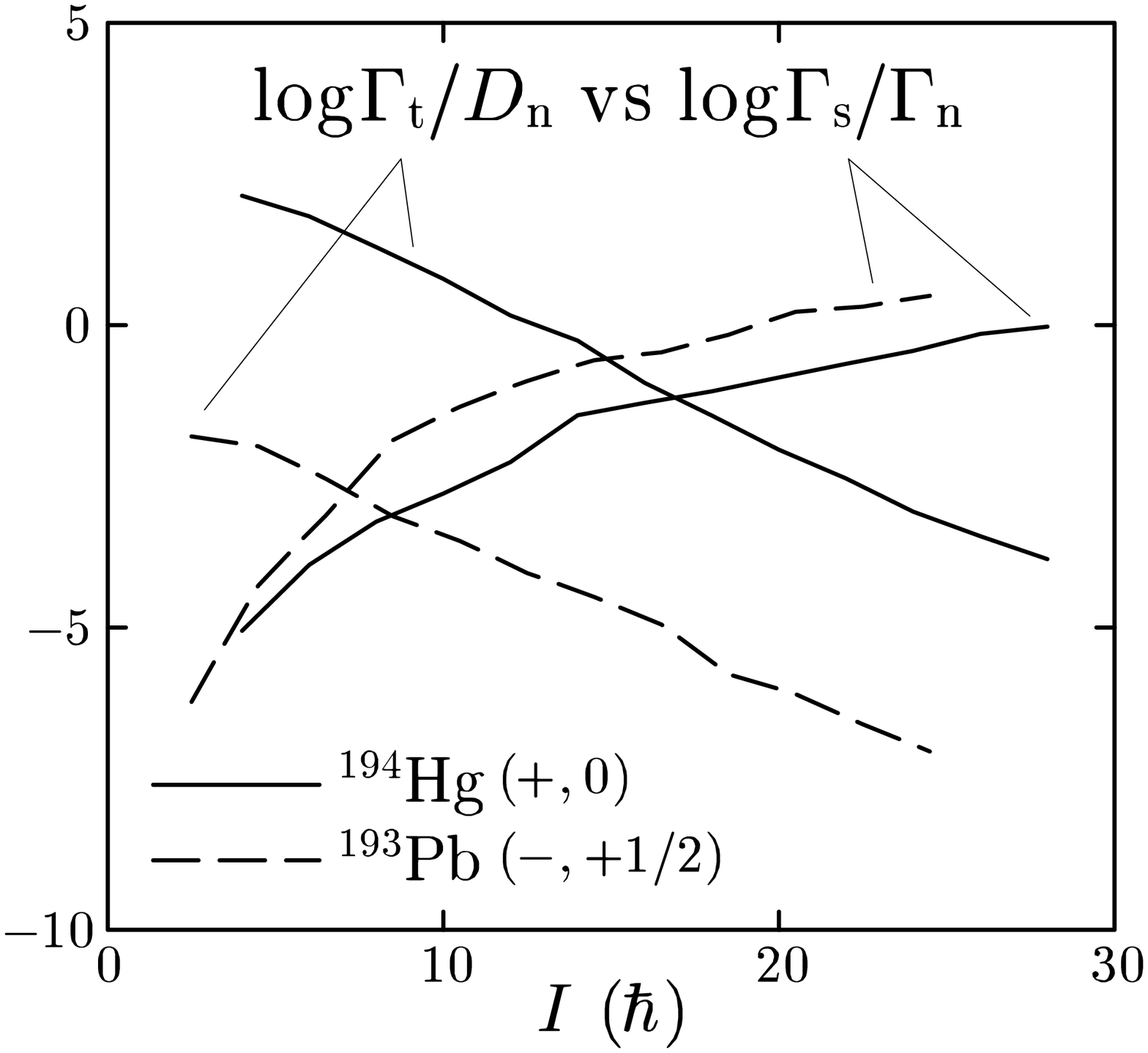}
}
\vspace{-10mm}
\caption{
 $\log_{10}(\Gammat/D_\rmn)$ versus $\log_{10}(\Gammas/\Gamman)$
 as functions of spin.}
\label{fig:GamDsnI}
\end{minipage}
\hspace{\fill}
\begin{minipage}[t]{76mm}
\centerline{
\epsfxsize=68mm\epsffile{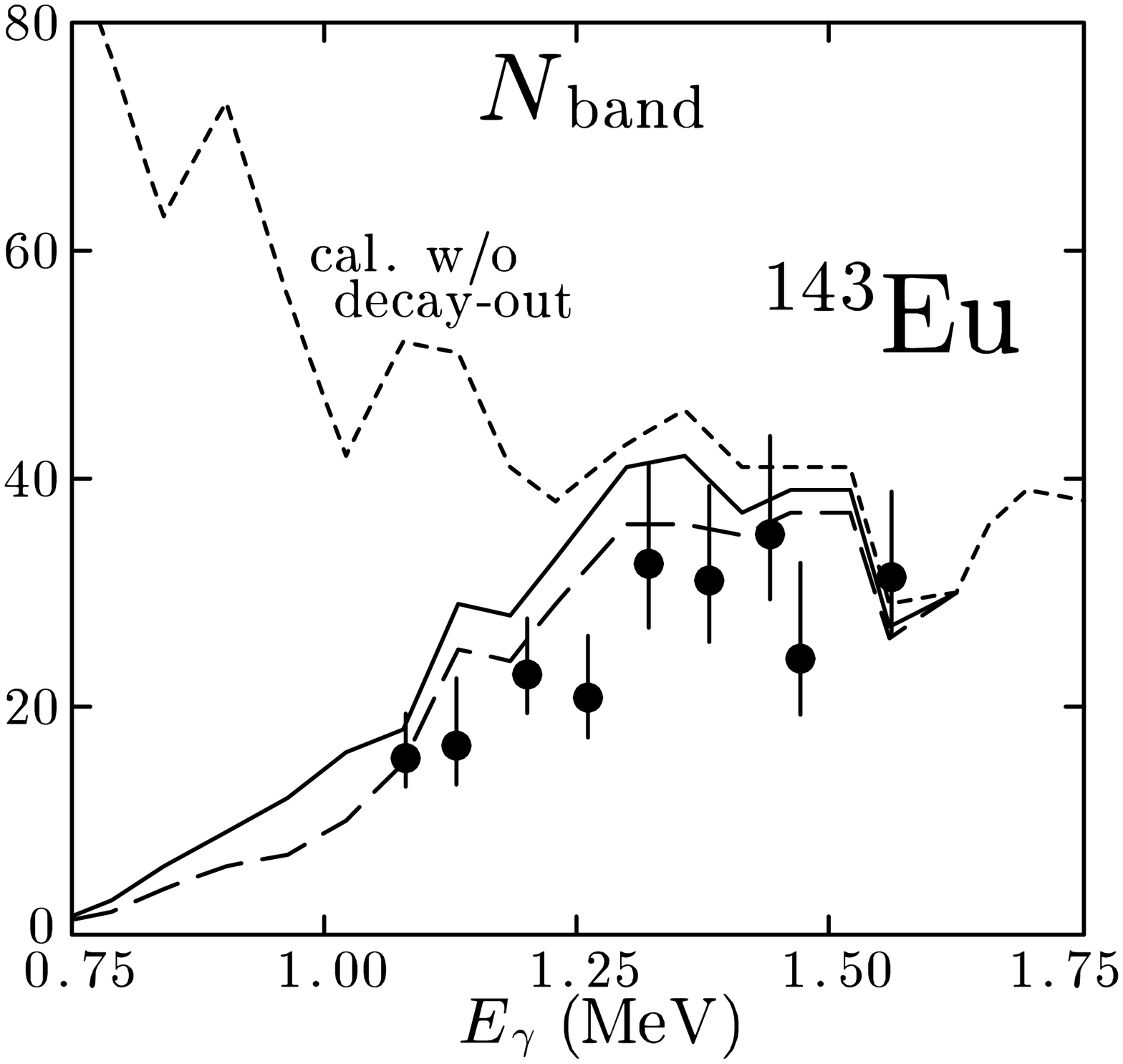}
}
\vspace{-10mm}
\caption{
 Comparison of calculated and measured
 number of bands for $^{143}$Eu.}
\label{fig:Nband}
\end{minipage}
\vspace{-5mm}
\end{figure}

In order to see how the decay-out spin is determined by
the four relevant quantities, we note that
$\Nout \approx \sqrt{(\pi/2)(\Gammat/\Dn)/(\Gammas/\Gamman)}$ holds
in realistic situations~\cite{VBD90,SDVB93}.
Two quantities, $\Gammat/\Dn$ and $\Gammas/\Gamman$,
are plotted as functions of spin in Fig.~\ref{fig:GamDsnI}.
$\Nout$ takes appreciable value when $\Gammat/\Dn>\Gammas/\Gamman$,
i.e. two curves crosses in Fig.~\ref{fig:GamDsnI},
and $\Iout$ is roughly determined by their crossing point.
Moreover, the decay-out occurs more rapidly
as the crossing angle between the two curves gets larger.
$\log_{10}(\Gammas/\Gamman)$ decreases more rapidly
in the low-spin region because of the $E_\gamma^5$ behaviour
of the rotational E2 transitions of SD bands, and therefore
the decay-out tends to be more rapid when $\Iout$ is lower;
this behaviour is well observed in Figs.~\ref{fig:A150} and~\ref{fig:A190}.
$\Gammas/\Gamman$ takes similar values for SD bands within
the $A \approx 150$ or 190 region,
while $\Gammat/\Dn$ takes variety of values
mainly due to the tunneling probability,
which reflects the different potential landscapes
depending on configurations and nuclei.
This is the main reason why the range of calculated $\Iout$
spreads rather broadly in contradiction to experimentally observed trend.
In contrast, observed hindrance factors for the decay of $K$-isomers
take broad range of values in accordance with
the results of similar calculations of tunneling~\cite{NSS96}.
The observed trend that $\Iout$ falls in a rather narrow range
in both the $A \approx 150$ and 190 regions
is not easy to understand.

Finally, we have recently developed a microscopic model
of thermally excited SD states that describes both
the barrier penetration leading to the decay-out to the ND states,
and the rotational damping causing fragmentation of rotational
E2 transitions~\cite{YMS00}.
As an example, $\Nband$, the number of rotational SD bands
in the quasi-continuum $\gamma$-ray region for $^{143}$Eu
is shown in Fig.~\ref{fig:Nband}, where solid (dashed) curve
is a result using Eq.~(1c) (Eq.~(1b)).
Inclusion of the barrier penetration reduces $\Nband$ dramatically
in the low-spin region and the calculation nicely agrees
with experimental data~\cite{Silv00},
which clearly shows that our model is capable for the unified
understanding of feeding and decay of the SD bands.

\section{SUMMARY}

Systematic comparison of the calculated and observed relative intensities
have been performed, and it is found that basic features of
the decay-out of SD bands are well understood
within our theoretical framework.
However, the decay-out spin is not precisely reproduced
in the calculation in individual cases.
Especially an average value of $\Iout$ for the $A \approx 190$ nuclei
is somewhat overestimated, and the range of calculated $\Iout$'s seems
too wide in both the $A \approx 150$ and 190 regions.
It should, however, be noticed that
the average value of calculated $\Iout$ in the $A \approx 190$ region
is much smaller than that in the $A \approx 150$ region by about 10$\hbar$;
the difference in the experiment is even more, about 15$\hbar$, though.
As a possible source of improvement a simple scaling of
the mass tensor is considered.
We believe that microscopic calculations of mass tensor
based on counting the number of level crossings has to be done
in order to include detailed nuclear structure effects in each SD band.
For the detailed comparison, the reproduction of excitation
energies of the SD bands is very important,
which is not enough in the present calculations;
for example, that of $^{194}$Pb in Table~\ref{tab:Gamma}
is about 1 MeV higher than the experimentally deduced value.
For this purpose, the improvement of the potential energy surface
may also be necessary, e.g. by making use of an
another mean-field potential like Woods-Saxon or the one obtained
by Skyrme Hartree-Fock method.  These are remaining future problems.

As an another possible source of improvement,
modification of the compound mixing model~\cite{VBD90} may be considered.
At higher spins where the excitation energy of SD bands is not so high,
the ND states to which the SD band couples are not chaotic,
and then strong reduction of mixing amplitudes
may be expected~\cite{Aberg99}.
This reduction leads to a delay of decay-out,
which hopefully helps to decrease $\Iout$
especially in the $A \approx 190$.
There are, however, some ambiguities of the amount of reduction,
and a precise modeling of this effect is necessary
to draw a definite conclusion.
It is also an interesting future problem.

\vspace{5mm}

This work is supported in part by the Grant-in-Aid for
Scientific Research from the Japan Ministry of Education,
Science and Culture (No. 12640281).

\end{document}